\documentclass[a4paper,12pt]{article}
\pdfoutput=1

\usepackage{amsmath,amssymb,amsfonts}% Typical maths resource packages
\allowdisplaybreaks
\usepackage{graphicx}% Packages to allow inclusion of graphics
\usepackage{color}% For creating coloured text and background
\makeatletter
\@addtoreset{equation}{section}
\renewcommand{\theequation}{\thesection.\@arabic\c@equation}
\makeatother
\usepackage{hyperref}% For creating hyperlinks in cross references
\usepackage{cite}
\usepackage{caption}
\definecolor{red}{rgb}{1,0,0}% Standard colours red, green, blue
\definecolor{green}{rgb}{0,1,0}
\definecolor{blue}{rgb}{0,0,1}
\definecolor{darkblue}{rgb}{0,0,0.8}
\definecolor{darkred}{rgb}{0.6,0,0}
\definecolor{lightblue}{rgb}{.5,.5,1}
\definecolor{lightgray}{gray}{.87}% How you can define your own greys
\definecolor{Dark}{gray}{.20}
\definecolor{pink}{rgb}{.95,0.82,0.92}% How you can define your own colours
\definecolor{yellow}{rgb}{1,1,0}
\definecolor{lightyellow}{rgb}{1,1,.5}
\definecolor{purple}{rgb}{0.7,0,0.85}
\definecolor{darkgreen}{rgb}{0,0.25,0}
\definecolor{darkpurple}{rgb}{0.4,0,0.4}
\definecolor{orange}{rgb}{0.8,0.2,0.2}
\def \be {\begin{equation}}
\def \ee {\end{equation}}
\def \bea {\begin{align}}
\def \eea {\end{align}}
\def \nn {\nonumber}

\def \rr {\raise.35ex\hbox{\small $\prime$}\kern-.17em{\mbox{\large $\imath$}}}
\def \del {\partial}
\def \dels {\partial\kern-.5em / \kern.5em}
\def \As {{A\kern-.5em / \kern.5em}}
\def \Ds {D\kern-.7em / \kern.5em}

\def \a {\alpha}

\def \b {\beta}
\def \dag {\dagger}

\def \d {\delta}
\def \eps {\epsilon}

\def \lam {\lambda}
\def \Lam {\Lambda}

\def \om {\omega}
\def \Om {\Omega}

\def \Th {\Theta}

\newcommand{\detail}[1]{}

\newcommand{\hide}[1]{}

\newcommand{\vtwo}[1]{{\color{darkblue} #1}}

\newcommand{\explanation}[1]{}

\begin{document}

\pagestyle{plain}

%\begin{CJK}{UTF8}{bsmi} 

\begin{titlepage}
\vspace*{-10mm}   
\baselineskip 10pt   
\begin{flushright}   
\begin{tabular}{r} 
%KUNS-nnn\\
%RIKEN-iTHEMS-Report-21
%April 1, 2018
\end{tabular}   
\end{flushright}   
\baselineskip 24pt   
\vglue 10mm

\begin{center}

\noindent
\textbf{\LARGE
Planckian Physics Comes Into Play\\
\vskip0.3em
At Planckian Distance From Horizon
}
\vskip20mm
\baselineskip 20pt

\renewcommand{\thefootnote}{\fnsymbol{footnote}}

{\large
Pei-Ming~Ho${}^{a,b}$
\footnote[1]{pmho@phys.ntu.edu.tw},
Hikaru~Kawai${}^{a,b}$
\footnote[2]{hikarukawai@phys.ntu.edu.tw},
Yuki~Yokokura$^c$
\footnote[3]{yuki.yokokura@riken.jp}
}

\renewcommand{\thefootnote}{\arabic{footnote}}

\vskip5mm

{\it  
${}^{a}$
Department of Physics and Center for Theoretical Physics, \\
National Taiwan University, Taipei 106, Taiwan,
R.O.C. 
\\
${}^{b}$
Physics Division, National Center for Theoretical Sciences, \\
Taipei 106, Taiwan,
R.O.C. 
\\
${}^{c}$
iTHEMS Program, RIKEN, Wako, Saitama 351-0198, Japan
}

\vskip 25mm
\begin{abstract}

In the background of a gravitational collapse,
we compute the transition amplitudes for
the creation of particles for distant observers
due to higher-derivative interactions
in addition to Hawking radiation.
The amplitudes grow exponentially with time
and become of order $1$
when the collapsing matter is about a Planck length outside the horizon.
As a result,
the effective theory breaks down at the scrambling time,
invalidating its prediction of Hawking radiation.
Planckian physics comes into play to decide the fate of black-hole evaporation.

\end{abstract}
\end{center}

\end{titlepage}

\pagestyle{plain}

\baselineskip 18pt

\setcounter{page}{1}
\setcounter{footnote}{0}
\setcounter{section}{0}

%\tableofcontents

\newpage

%%%%%%%%%%%%%%%%%%%%%%%%%%%%%%%%%%%%%%%%%%%%%%%%%%
%%%%%%%%%%%%%%%%%%%%%%%%%%%%%%%%%%%%%%%%%%%%%%%%%%
%%%%%%%%%%%%%%%%%%%%%%%%%%%%%%%%%%%%%%%%%%%%%%%%%%

\section{Introduction}\label{introduction}

The breakdown of the low-energy effective theory is anticipated
\cite{Mathur:2009hf,firewall}
for resolving the information loss paradox \cite{Hawking:1976ra}.
But how does the effective theory break down?
%\detail{Classically,
%no instability is found for Schwarzschild black holes \cite{Johnson:2018yci},
%Kerr black holes \cite{Dafermos:2014cua},
%or extremal black holes \cite{Hadar:2017ven}.}
Assuming string theory,
it is argued that the effective theory breaks down
due to longitudinal string-stretching effects \cite{Dodelson:2015toa}.
On the other hand,
there should be clear signals for the breakdown of an effective theory
even when we do not know anything about the UV theory.

In this work,
we examine the validity of the low-energy effective theory
for black-hole physics by
carrying out standard QFT calculations
under the assumption that the horizon is initially uneventful,
as in the conventional model of black holes.
Without proposing new physical principles,
the only novelty is that we take into consideration
the effect of higher-derivative interactions,
which are ubiquitous in generic low-energy effective theories
\cite{Weinberg:1995mt}.

Non-renormalizable contributions to a physical quantity
(e.g. a scattering amplitude) are typically suppressed by powers of $E^2/M_p^2$,
where $E$ is the characteristic energy of the physical process
and the Planck mass $M_p$ is by assumption the cutoff energy of the effective theory.
Hence,
non-renormablizable interactions have been ignored in most studies of black-hole physics
%\vtwo{
(except that its potential importance was mentioned in Ref.\cite{Iso:2008sq}).
%}
But they also play the role of an indicator of the validity of the effective theory.
The effective theory breaks down
whenever non-renormalizable contributions are large,
because it means $E \gtrsim M_p$,
and a UV theory is needed.

In particular,
higher-derivative (non-renormalizable) interactions play an important role
in black-hole physics for the following reasons.
First,
due to the trans-Planckian problem \cite{trans-Planckian-1,trans-Planckian-2}
of Hawking radiation \cite{Hawking:1974rv,Hawking:1974sw,Hawking:1976de},
ultra-energetic quantum modes must be included in the effective theory.
The energy scale $E$ for a process involving these modes can be large.
Secondly,
while the equivalence principle is what prevents the horizon from becoming eventful,
higher-derivative interactions have the potential to violate the equivalence principle
\cite{Lafrance:1994in}.

In this paper,
we will consider a spherical thin null shell in gravitational collapse
and compute in this background geometry the transition amplitudes 
for the following two physical processes:
(1) the creation of outgoing particles
from the Unruh vacuum due to
higher-derivative couplings to the background curvature,
\footnote{
This particle creation is independent of Hawking radiation,
which appears even for free field theories.
}
and 
(2) the decay of an ingoing particle 
into an ingoing one and two outgoing ones
through 4-point higher-derivative interactions.
The outgoing particles in both amplitudes are defined from the viewpoint of distant observers,
as we use asymptotic states in the calculation of scattering amplitudes.
In particular,
we consider outgoing particles with energies $\sim \mathcal{O}(1/a)$
because states including them must exist
in order for the effective theory to predict Hawking radiation. 
(Here,
$a \equiv 2G_N M$ is the Schwarzschild radius,
where $G_N$ is the Newton constant
and $M$ is the black hole mass.)

It turns out that these amplitudes grow exponentially with time
and become $\mathcal{O}(1)$
at the scrambling time
\footnote{
The scrambling time $\Delta t \sim \mathcal{O}(a \log(aM_p))$
was initially introduced in a different context \cite{Sekino:2008he}.
}
after the collapsing matter reaches $r - a \sim \mathcal{O}(a)$.
At this time,
the areal radius of the matter is close to $r = a + \mathcal{O}(\ell_p^2/a)$,
where $\ell_p \equiv 1/M_p$ is the Planck length,
and the proper distance between the matter and the horizon is of $\mathcal{O}(\ell_p)$.
%when the collapsing matter is slightly outside the horizon
%with an areal radius $R - a \sim \mathcal{O}(\ell_p^2/a)$
%(at a proper distance of $\mathcal{O}(\ell_p)$ from the horizon),
%where $\ell_p \equiv 1/M_p$ is the Planck length.
As a result,
the effective theory becomes invalid
for physics at the energy scale of Hawking radiation.

%%%
The key point behind the particle creations from vacuum is the uncertainty principle
for the non-commutativity of the two momentum operators
defined with respect to distant observers and freely falling observers.
Because of this uncertainty principle,
momentum conservation for freely falling observers
hardly restricts the creation of outgoing particles for distant observers.
%%%
\detail{
Such a claim about particle creation may sound incomprehensible at first sight.
As the spacetime curvature is small for a large black hole,
it is almost the same as Minkowski space in a local freely falling frame.
The amplitude for particle creation should be negligible
as it violates energy conservation,
instead of growing exponentially with time.
We will resolve this apparent contradiction.
It has to do with the uncertainty relation between two momentum operators
defined with respect to different time coordinates.
}

For essentially the same reason,
the amplitude for a freely falling particle
to radiate outgoing particles to distant observers
due to higher-derivative interactions
also grows exponentially with time.
It may provide a mechanism to transfer information
from the infalling particle to outgoing radiation.

The result that
the effective theory breaks down around the scrambling time
for the energy scale of Hawking radiation means that
a UV theory is needed to decide 
whether Hawking radiation continues
and whether the horizon remains uneventful.
Regardless of what happens in the UV theory,
however,
the information loss paradox is no longer a paradox for the effective theory.

The plan of this paper is as follows.
We review in Sec.\ref{background} necessary elements about
quantum scalar fields in the spacetime background of a matter shell in gravitational collapse.
The trans-Planckian problem is reviewed in Sec.\ref{TPP},
and the issue of energy conservation in a local freely falling frame
and the uncertainty principle are explained in Sec.\ref{FFF}.
The amplitude for the creation of particles due to higher-derivative couplings
to the background Ricci tensor is calculated in Sec.\ref{HDPC}.
%The characteristic energy $E$ is deduced from
%the explicit expressions of the amplitudes in Sec.\ref{HR=F}.
In Sec.\ref{TAIP},
we compute the amplitude for the radiation of freely falling particles
due to higher-derivative interactions.
Finally,
we make comments in Sec.\ref{Conclusion}
on the generalization of the calculations in this paper
and different scenarios for black-hole evaporation.

%%%%%%%%%%%%%%%%%%%%%%%%%%%%%%%%%%%%%%%%%%%%%%%%%%
%%%%%%%%%%%%%%%%%%%%%%%%%%%%%%%%%%%%%%%%%%%%%%%%%%
%%%%%%%%%%%%%%%%%%%%%%%%%%%%%%%%%%%%%%%%%%%%%%%%%%

\section{Review of scalar field in Schwarzschild background}
\label{background}

We review in this section
the quantum physics for a black hole with spherical symmetry.
\footnote{
For a more comprehensive review,
see Ref.\cite{Brout:1995rd}.
}
The Schwarzschild metric outside the collapsing matter is
\begin{align}
ds^2 
&= - \left(1 - \frac{a}{r}\right) du dv + r^2(u, v) d\Omega^2.
\label{Schwarzschild-metric}
\end{align}
%where the Schwarzschild radius $a = 2G_N M$
%for a black hole of mass $M$.
%($G_N$ is the Newton constant.)
The areal radius $r(u, v)$ is implicitly defined by the equation
\begin{align}
\frac{v - u}{2} &= r + a \log\left(\frac{r}{a} - 1\right).
\label{r-tortoise}
\end{align}
The light-cone coordinates $(u, v)$ coincide with
the retarded and advanced light-cone coordinates
in the asymptotically flat region.
For simplicity,
we consider the gravitational collapse of a null thin shell at $v = v_s$,
so the metric \eqref{Schwarzschild-metric} holds only for $v \geq v_s$.

In the near-horizon region where $0 < r - a \ll a$,
the Schwarzschild metric is approximately
\be
ds^2 \simeq - dU dV + r^2 d\Omega^2,
\label{metric-NHR}
\ee
where
\begin{align}
U(u) &\simeq - 2a e^{- \frac{u}{2a}},
\label{U-def}
\\
V(v) &\simeq 2a e^{\frac{v}{2a}}
\label{V-def}
\end{align}
are (up to a constant factor) the Kruskal coordinates.
These are the coordinates suitable for freely falling observers comoving with the collapsing matter.
The red-shift factor between the retarded light-cone coordinates $u$ and $U$,
\begin{align}
\frac{dU}{du} &\simeq e^{- \frac{u}{2a}},
\label{dUdu}
\end{align}
decays exponentially with $u$.
The future horizon is located at $u = \infty$ (or equivalently, $U = 0$).

The metric inside the collapsing shell is the Minkowski metric:
\begin{align}
ds^2 &= - dUdV + R^2(U, V) d\Omega^2,
\label{Mink}
\end{align}
where
\begin{align}
R(U, V) \equiv \frac{V - U}{2}.
\label{R}
\end{align}
We have used the same symbols $(U, V)$ for these coordinates
because they can be identified with the Kruskal coordinates
\eqref{U-def} and \eqref{V-def} across the collapsing shell
in the near-horizon region.

Ignoring the back-reaction of the vacuum energy-momentum tensor,
the energy-momentum tensor of the null shell has the only non-vanishing component
\begin{align}
T_{VV} = \frac{M}{4\pi R^2(U, V)} \, \d(V - V_s),
\label{TVV}
\end{align}
where $V_s = 2a \exp(v_s/2a)$ is the $V$-coordinate of the null thin shell.
At the horizon,
$R(0, V_s) = a$,
so eq.\eqref{R} implies that,
in our convention,
$v_s = 0$ and
\begin{align}
V_s = 2a.
\label{Vs}
\end{align}

For a very large black hole,
the curvature is small.
%%% p1
For the spacetime region
and the frequencies to be considered below,
the $s$-wave of a quantum field $\phi$ is well approximated by
%%% 1p
\begin{align}
\phi &\simeq \int_0^{\infty} \frac{d\Om}{4\pi\sqrt{\Om} R(U,V)}
\left(
a_{\Om} e^{- i\Om U} + a^{\dag}_{\Om} e^{i\Om U}
+ \tilde{a}_{\Om} e^{- i\Om V} + \tilde{a}^{\dag}_{\Om} e^{i\Om V}
\right)
\nn \\
&= \int_0^{\infty} \frac{d\om}{4\pi\sqrt{\om} R(U(u),V(v))}
\left(
b_{\om} e^{- i\om u} + b^{\dag}_{\om} e^{i\om u}
+ \tilde{b}_{\om} e^{- i\om v} + \tilde{b}^{\dag}_{\om} e^{i\om v}
\right),
\label{phi-modes}
\end{align}
where the creation-annihilation operators satisfy
the canonical commutation relations:
\begin{align}
&[a_{\Om}, a^{\dag}_{\Om'}] = \d(\Om - \Om'),
\quad
&[\tilde{a}_{\Om}, \tilde{a}^{\dag}_{\Om'}] = \d(\Om - \Om'),
\label{a-CR}
\\
&[b_{\om}, b^{\dag}_{\om'}] = \d(\om - \om'),
\quad
&[\tilde{b}_{\om}, \tilde{b}^{\dag}_{\om'}] = \d(\om - \om').
\label{b-CR}
\end{align}
The Unruh vacuum $| 0 \rangle$ can be defined by
\be
a_{\Om}|0\rangle = 
\tilde{a}_{\Om}|0\rangle = 0.
\ee
It can be identified with the Minkowski vacuum
of the infinite past.

The creation-annihilation operators $(a^{\dag}_{\Om}, a_{\Om})$ are 
defined with respect to the Kruskal coordinate $U$,
and $(b^{\dag}_{\om}, b_{\om})$ those
defined with respect to fiducial observers (or distant observers).
They are related to each other via the Bogoliubov transformation
\begin{align}
b_{\om} &= \int_0^{\infty} d\Om \, 
\left(\a_{\om\Om} a_{\Om} + \b_{\om\Om} a^{\dag}_{\Om}\right),
\label{Bogo-1}
\\
b^{\dag}_{\om} &= \int_0^{\infty} d\Om \, 
\left(\b^{\ast}_{\om\Om} a_{\Om} + \a^{\ast}_{\om\Om} a^{\dag}_{\Om}\right),
\label{Bogo-2}
\end{align}
where
\begin{align}
\a_{\om\Om} &\equiv \frac{1}{2\pi} \sqrt{\frac{\om}{\Om}}
\int_{-\infty}^{\infty} du \, e^{i\om u - i\Om U(u)}
\simeq \frac{a}{\pi} \sqrt{\frac{\om}{\Om}} (2a\Om)^{i2a\om}
e^{\pi a\om} \Gamma(-i 2a\om),
\label{alpha}
\\
\b_{\om\Om} &\equiv \frac{1}{2\pi} \sqrt{\frac{\om}{\Om}}
\int_{-\infty}^{\infty} du \, e^{i\om u + i\Om U(u)}
\simeq \frac{a}{\pi} \sqrt{\frac{\om}{\Om}} (2a\Om)^{i2a\om}
e^{- \pi a\om} \Gamma(-i 2a\om),
\label{beta}
\end{align}
with $U(u)$ given by eq.\eqref{U-def} in the near-horizon region.

A proper description of Hawking radiation requires the use of wave packets,
as Hawking noticed in his original paper \cite{Hawking:1974sw},
so that one can talk about the radiation at different times during the gravitational collapse.
Furthermore,
a physical particle is always detected within a finite region of space and time,
as opposed to being a uniform spacetime-filling plane wave.

A wave packet allows us to talk about a particle of an approximate energy $\om_0$
localized within a neighborhood of size $\Delta u$ centered at a certain point $u_0$.
The uncertainty $\Delta\om$ in the frequency $\om_0$
and the size $\Delta u$ obey the uncertainty relation $\Delta u\Delta\om \geq 1$.

For a well-localized wave packet around $u_0$
with a given frequency distribution $f_{\om_0}(\om)$,
the corresponding creation operator is
\footnote{
\label{HWP}
Hawking considered wave packets with the frequency distribution
$
f_{\om_0}(\om) =
\frac{1}{\sqrt{\eps}} (\Th(\om - \om_0) - \Th(\om - \om_0 - \eps))
\label{Hawking-WP}
$
in his original work \cite{Hawking:1974sw}.
}
\begin{align}
b^{\dag}_{(\om_0, u_0)} &= 
\int_0^{\infty} d\om \, f_{\om_0}(\om) e^{i \om u_0} b^{\dag}_{\om},
\label{bdag-wp}
\end{align}
%and similarly for the annihilation operators $b_{(\om_0, u_0)}$,
where $f_{\om_0}(\om)$ is assumed to be normalized:
\begin{align}
\int_0^{\infty} d\om \, f^{\ast}_{\om_0}(\om) f_{\om_0}(\om) = 1,
\label{f-norm}
\end{align}
so that $[b_{(\om_0,u_0)}, b^{\dag}_{(\om_0,u_0)}] = 1$.

As a simple example,
we will use the Gaussian wave packet
with the frequency distribution
\be
f_{\om_0}(\om) = c_{\om_0} \, e^{- \frac{(\om - \om_0)^2}{2(\Delta\om)^2}},
%\qquad
%\vthree{
%c_{\om_0} \equiv \left[\frac{1}{2}\sqrt{\pi} (\Delta\om)\right]^{-1/2},
%}
\label{f-Gaussian}
\ee
%\vtwo{
where $c_{\om_0}$ is determined by the normalization condition \eqref{f-norm} 
(up to a phase factor):
\be
c_{\om_0} = \left[\int_0^{\infty} d\om \, e^{- \frac{(\om - \om_0)^2}{(\Delta\om)^2}}\right]^{-1/2}
\simeq \left[\frac{\sqrt{\pi} (\Delta\om)}{2}\right]^{-1/2}
\quad \mbox{for $\om_0 \gg (\Delta\om)$}.
\ee
%}

%%% p1
Hawking radiation can be described in terms of
$\langle 0 | b^{\dag}_{(\om_0, u_0)}b_{(\om_0, u_0)} | 0 \rangle$
for a given wave packet,
which is proportional to the Planck distribution \cite{Hawking:1974sw}.
\hide{
\begin{align}
\langle 0 | b^{\dag}_{(\om_0, u_0)} b_{(\om_0, u_0)} | 0 \rangle
&\simeq
\frac{\Gamma_{(\om_0, u_0)}}{e^{4\pi a\om_0} - 1},
\end{align}
where $\Gamma_{(\om_0, u_0)}$ is a time-dependent factor
that describes how Hawking radiation gradually appears
in the dynamical process of gravitational collapse.
At late times $(u_0 \rightarrow \infty)$, 
$\Gamma_{(\om_0, u_0)} \rightarrow \Delta u/2\pi$ \cite{Brout:1995rd}.
}

\hide{
[Remove this later.]
The factor $\Gamma_{(\om_0, u_0)}$ must be time-dependent because,
at least for $u_0 \rightarrow - \infty$,
Hawking radiation must vanish.
This is why the wave packet is needed ---
to show
``how Hawking radiation gradually appears
in the dynamical process''.
}
%%% 1p

%%%%%%%%%%%%%%%%%%%%%%%%%%%%%%%%%%%%%%%%%%%%%%%%%%
%%%%%%%%%%%%%%%%%%%%%%%%%%%%%%%%%%%%%%%%%%%%%%%%%%
%%%%%%%%%%%%%%%%%%%%%%%%%%%%%%%%%%%%%%%%%%%%%%%%%%

\section{Trans-Planckian problem}
\label{TPP}

Naively,
\footnote{
Strictly speaking,
as $dU/du$ \eqref{dUdu} changes exponentially with time,
a constant frequency $\om$ is mapped to a wide range of frequencies $\Om$,
instead of a unique frequency \eqref{dom-omp}.
A more precise description of this mapping between $\om$ and $\Om$
in terms of wave packets
will be given in Sec.\ref{FFF}.
}
the frequency $\om$ defined with respect to $u$
corresponds to the frequency $\Om$ defined with respect to $U$
through the red-shift factor $dU/du$ as
\begin{align}
\Om \sim \left(\frac{dU}{du}\right)^{-1} \om.
\label{dom-omp}
\end{align}
Since Hawking radiation is dominated by the frequencies $\om \sim \mathcal{O}(1/a)$,
if the low-energy effective theory is only valid for
frequencies $\om$ below a certain cutoff $\Lam_{\om}$,
we must have
\be
\Lam_{\om} > \frac{1}{a}
\label{lam>1/a}
\ee
for Hawking radiation to be a reliable prediction.
Through %eq.\eqref{dUdu} and 
eq.\eqref{dom-omp},
one expects that
the cutoff $\Lam_{\om}$ on $\om$ is translated to a cutoff $\Lam_{\Om}$ on $\Om$ with
\begin{align}
\Lam_{\Om} > \left(\frac{dU}{du}\right)^{-1} \frac{1}{a}.
\label{dom-omp-1}
\end{align}

\detail{
In Appendix \ref{HR},
we substantiate this heuristic argument by
a more rigorous derivation of eq.\eqref{dom-omp-1}
as the necessary condition for Hawking radiation
to be a valid prediction of the low-energy effective theory.
}

Over the evaporation of a time scale $\sim \mathcal{O}(a^3/\ell_p^2)$,
say, for Hawking radiation to persist till the Page time,
eqs.\eqref{dUdu} and \eqref{dom-omp-1} lead to
\begin{align}
\Lam_{\Om} > \mathcal{O}\left(\frac{1}{a} \, e^{a^2/\ell_p^2}\right).
\label{large-Lam}
\end{align}
Thus,
Hawking radiation demands the validity of the effective theory
for extremely high frequencies $\Om$.
This is the trans-Planckian problem 
\cite{trans-Planckian-1,trans-Planckian-2}
of Hawking radiation.

Some have argued that this trans-Planckian problem can be dismissed.
First of all,
it was shown that Hawking radiation persists
\cite{trans-Planckian-3} in free-field theories 
with modified dispersion relations
for which the energy is always smaller than the Planck mass $M_p$.
\footnote{
See Ref.\cite{trans-Planckian-4} for reservations about
the implications of these works.
}
Renormalizable interactions were also studied
\cite{Leahy:1983vb,Helfer:2005wz,Frasca:2014gua}
and no significant effect on Hawking radiation was found.

A natural next step is to study the effect of
higher-derivative non-renormalizable interactions on Hawking radiation,
as they amplify the contributions of high-frequency modes.
Moreover,
the effective theory breaks down
whenever their contributions are large.

%Recently,
The potential importance of
higher-derivative non-renormalizable interactions were suggested in Ref. \cite{Iso:2008sq},
%\cite{Ho:2020cbf,Ho:2020cvn}.
and it was recently found
%It was found 
\cite{Ho:2020cbf,Ho:2020cvn} that,
from the viewpoint of distant observers,
higher-derivative interactions induce large amplitudes for the creation of particles.
However,
without using the notion of wave packets,
the calculation was not rigorous,
and there was an misinterpretation of the result
in Refs.\cite{Ho:2020cbf,Ho:2020cvn}.
We will give a proper interpretation in Sec.\ref{FFF}
on the large amplitudes,
and a cleaner and more rigorous derivation
of a similar result in Sec.\ref{HDPC}.

Another argument against the trans-Planckian problem is that
the frequency $\Om$ for outgoing particles is not a local Lorentz invariant.
It can be arbitrarily large or small after a local Lorentz boost:
$U \rightarrow \lam U$ ($\lam \in \mathbb{R}_+$),
which corresponds to $\Om \rightarrow \lam^{-1}\Om$.
Since the cutoff energy of an effective theory should be locally Lorentz invariant
\footnote{
This is why we can apply the standard model
to ultra-high-energy cosmic rays.
},
the cutoff $M_p$ of the effective theory 
does not imply a cutoff like $M_p > \Om$.

This argument can be refuted by 
constructing an invariant energy $E$ from $\Om$.
As $\Om$ is the $U$-component of the momentum
(to be denoted by $P_U$) of a quantum mode,
we only need the $V$-component of some momentum
(to be denoted by $P_V$) to define an invariant.
In addition to the collapsing matter which clearly has a non-zero $P_V$,
there is also an ingoing negative vacuum energy flux in the near-horizon region
\cite{Davies:1976ei,Christensen:1977jc}.
In either case,
we expect the background to have a $P_V$-component no less than $P_V \sim 1/a$.
Hence,
an invariant characteristic energy can then be defined as
\begin{align}
E^2 \equiv P_U P_V \sim \frac{\Om}{a}.
\label{E-def}
\end{align}
When we fix the outgoing momentum to be $\om \sim \mathcal{O}(1/a)$,
we plug the corresponding $\Om$ \eqref{dom-omp}
into eq.\eqref{E-def} and find the invariant energy scale
\be
E^2 \sim \frac{1}{a^2} e^{\frac{u}{2a}}.
\label{E>}
\ee

With the enhancement of higher-derivatives,
the amplitude for particle creation,
as a function of such an invariant energy $E$,
is expected to be of the form
\begin{align}
\mbox{\em Amplidtude}
\sim \left(\frac{E^2}{M_p^2}\right)^m
\sim \left(\frac{\ell_p^2}{a^2} \, e^{\frac{u}{2a}}\right)^m,
\label{ampl-exp}
\end{align}
where we expect a larger $m$ for a higher-derivative interaction of higher order.

This argument might be considered unreliable because
the energy $E$ \eqref{E-def} involves the frequency $\Om$ of a virtual particle,
instead of a real one.
But we will show explicitly in Sec.\ref{HDPC} that
the amplitudes for creating real particles for distant observers
due to a class of higher-derivative interactions
indeed grow exponentially with time precisely as eq.\eqref{ampl-exp} with $m > 0$.
As a result,
at the scrambling time
\be
u \sim 2a \log(a^2/\ell_p^2),
\ee
the amplitudes become of order $1$,
and the effective theory breaks down.
We will consider the amplitudes for other processes in Sec.\ref{TAIP}
that also grow exponentially with time.
The trans-Planckian problem
does lead to the breakdown of the effective theory.

An important point to emphasize here is that
the created particles with large amplitudes
are defined with respect to distant observers,
rather than freely falling observers.
More precisely, 
these particles are defined by localized wave packets with approximate frequencies $\om$.
This will be the crucial point in resolving a seemingly paradoxical question
about momentum conservation in the freely falling frame in the next section.
Another potential concern the reader may have is 
how a large amplitude \eqref{ampl-exp} for non-renormalizable interactions
is compatible with the ``nice-slice argument'' \cite{Lowe:1995ac,Polchinski:1995ta}.
We will also answer this question in the next section.

%%%%%%%%%%%%%%%%%%%%%%%%%%%%%%%%%%%%%%%%%%%%%%%%%%
%%%%%%%%%%%%%%%%%%%%%%%%%%%%%%%%%%%%%%%%%%%%%%%%%%
%%%%%%%%%%%%%%%%%%%%%%%%%%%%%%%%%%%%%%%%%%%%%%%%%%

\section{Energy conservation and uncertainty relation}
\label{FFF}

Naively, 
the detection of particles of $\om \sim 1/a$
on top of Hawking radiation
would imply the existence of trans-Planckian particles
with extremely large $\Om$ for freely falling observers.
This is obviously in conflict with momentum conservation
in the freely falling frame.
In this section,
we explain why large amplitudes for particle creation at distances
are not ruled out by momentum conservation.
As we will see below,
there are two key points for resolving this conflict.
One is to use wave packets to formulate conservation laws in a more precise way.
The other is the uncertainty relation for the two momentum operators
defined with respect to distant observers and freely falling observers.

First,
in the near-horizon region,
the translation of $U$ is an approximate symmetry of the metric \eqref{metric-NHR}.
But the translation of $u$
(which is equivalent to a scaling of $U$ without scaling $V$)
is not a symmetry.
\footnote{
The simultaneous translation of $u$ and $v$
%(which is equivalent to opposite simultaneous scaling of $U$ and $V$)
is a symmetry of the Schwarzschild metric,
but it is broken by the background of collapsing matter at $v = v_s$.
}
As a result,
the component $P_U$ of the momentum is approximately conserved
but $P_u$ is not.
Hence,
although Hawking radiation carries positive $P_u$,
it is allowed to emerge from the vacuum.
%Total energy is conserved only after the back-reaction is taken into consideration.)

The outgoing particles in Hawking radiation also do not
violate $P_U$-conservation
because Hawking particles are only virtual particles
for freely falling observers.
On the other hand,
the state with 1 outgoing particle on top of Hawking radiation for distant observers,
\be
b^{\dag}_{\om} | 0 \rangle 
= \int_0^{\infty} d\Om \, \a^{\ast}_{\om\Om} \, a^{\dag}_{\Om} | 0 \rangle,
\ee
is simply a superposition of 1-particle states $a^{\dag}_{\Om} | 0 \rangle$
for freely falling observers,
where we have used the Bogoliubov transformation \eqref{Bogo-2}.
Similarly, 
the state with 2 outgoing particles on top of Hawking radiation for distant observers
is a superposition of 2-particle states for freely falling observers because
\begin{align}
b^{\dag}_{\om} b^{\dag}_{\om'} | 0 \rangle
&=
\int_0^{\infty} d\Om \int_0^{\infty} d\Om' \,
\a^{\ast}_{\om\Om} \a^{\ast}_{\om' \Om'} a^{\dag}_{\Om} a^{\dag}_{\Om'} | 0 \rangle.
\label{2+0}
\end{align}
Here we have used the identity
\be
\int_0^{\infty} d\Om \, \b^{\ast}_{\om\Om} \a^{\ast}_{\om' \Om} = 0
\ee
for $\om, \om' > 0$,
which we can check using eqs.\eqref{alpha} and \eqref{beta}.
That is,
the particles created on top of Hawking radiation for distant observers
correspond to particles on top of the Unruh vacuum for freely falling observers.
They carry both positive $P_u$ and positive $P_U$.
Naively,
the creation of these particles is expected to violate $P_U$-conservation.

Furthermore,
even when the outgoing particles have their $P_u$'s as small as $\mathcal{O}(1/a)$,
the corresponding $P_U$-values are naively extremely large
due to the large blue-shift factor $(dU/du)^{-1}$ at late times
(see eqs.\eqref{dUdu} and \eqref{dom-omp}).
%Regardless of whether the interaction has higher-derivatives or not,
The creation of such high-energy particles should be
strongly prohibited by $P_U$-conservation.
How is it possible to have large amplitudes
(like eq.\eqref{ampl-exp} for $m > 0$)
for the creation of such particles?
We resolve this apparent conceptual problem here.

Let us first review how $P_U$-conservation constrains an amplitude.
When we calculate scattering amplitudes
involving outgoing particles created by $a^{\dag}_{\Om}$,
which are $P_U$-eigenstates,
the amplitude is basically an integral over the product of the wave functions
(or their complex conjugates)
\footnote{
Please refer to Secs.\ref{HDPC} and \ref{TAIP} for the explicit calculations.
}
\begin{align}
\langle 0 | %\nabla_U^n 
\phi(U, V) a^{\dag}_{\Om} | 0 \rangle
&=
\frac{%(-i\Om)^n
1}{4\pi\sqrt{\Om} R(U, V)} \, e^{- i \Om U}
\end{align}
for each outgoing particle.
Apart from additional factors of $\Om$
for the $U$-derivatives in the interaction term,
the amplitude for creating $k$ outgoing particles from the vacuum is proportional to the integral
\begin{align}
\int_{-\infty}^{\infty} \frac{dU}{R^k(U, V)} \, e^{- i (\Om_1 + \cdots + \Om_k) U}.
\label{45}
\end{align}
If $R(U, V)$ is $U$-independent,
the integral is proportional to the Dirac $\d$-function $\d(\sum_{i=1}^k\Om_i)$,
which imposes $P_U$-conservation.
For the consideration of particle creation from the vacuum,
all particles have positive frequencies $\Om_i > 0$,
so that $\d(\sum_{i=1}^k\Om_i) = 0$,
and the amplitude vanishes.
In general,
for a smooth function $R(U, V)$ of $U$ with the characteristic length scale $L$,
the integral \eqref{45} for $\sum_{i=1}^k\Om_i \gg 1/L$ is very small.

However,
$R(U, V)$ can be well-defined in an effective theory
only if it is much larger than the Planck length $\ell_p$.
The integral \eqref{45} is ill-defined for a gravitational collapse
in which $R$ goes to $0$ within finite $U$.
One may thus wish to compute the amplitude up to a certain moment $U = U_{\ast}$
(e.g. up to the event horizon at $U = 0$).
The integral \eqref{45} is then replaced by
\begin{align}
\int_{-\infty}^{U_{\ast}} \frac{dU}{R^k(U, V)} \, e^{- i (\Om_1 + \cdots + \Om_k) U},
\label{47}
\end{align}
which is non-zero even when all $\Om_i$'s are positive.
That is,
for any finite $U_{\ast}$, 
$P_U$-conservation can be violated.
The standard interpretation in quantum mechanics
is that the measurement at an exact moment $U=U_{\ast}$
introduces an uncertainty in the momentum $P_U$ through the uncertainty relation
$\Delta U \Delta P_U \gtrsim 1$.

In a low-energy effective theory,
it is unphysical to introduce an instant $U_{\ast}$ with perfect precision.
We can avoid this problem using the notion of wave packets.
Consider a particle $a^{\dag}_{(\Om_0, U_0)}| 0 \rangle$,
where
\be
a^{\dag}_{(\Om_0, U_0)} \equiv
\int_0^{\infty} d\Om \, f_{\Om_0}(\Om) \, e^{i\Om_0 U_0} \, a^{\dag}_{\Om},
\ee
with the Gaussian wave packet \eqref{f-Gaussian}
(with $\Delta\om$ replaced by $\Delta\Om$).
Its wave function is
\begin{align}
\langle 0 | \phi(U, V) a^{\dag}_{(\Om_0, U_0)} | 0 \rangle
&\sim
\frac{(\Delta\Om)^{1/2}}{4\pi^{3/4}\Om_0^{1/2} R(U, V)} \,
e^{- \frac{(\Delta\Om)^2(U - U_0)^2}{2}} \,
e^{-i\Om_0(U - U_0)}.
\end{align}
(We shall not assume that $\Om_0 \gg \Delta\Om$ for the discussion below.
But even when $\Om_0 \ll \Delta\Om$,
there is only a factor of $2$ difference
in the approximation above.)

The amplitude for the creation of, say, two particles
in the state $a^{\dag}_{(\Om_1, U_1)} a^{\dag}_{(\Om_2, U_2)} | 0 \rangle$
would then be independent of the choice of $U_{\ast}$
as long as their wave packets reside well within the range $(- \infty, U_{\ast})$.
One may then replace the range $(- \infty, U_{\ast})$ by $(- \infty, \infty)$
in the integral of the wave functions.
For example,
for $k = 2$,
the amplitude \eqref{47} is replaced by
\begin{align}
&\int_{- \infty}^{\infty} dU \, 
\langle 0 | \phi(U, V) a^{\dag}_{(\Om_1, U_1)} | 0 \rangle
\langle 0 | \phi(U, V) a^{\dag}_{(\Om_2, U_2)} | 0 \rangle
\nn \\
&\simeq
\frac{e^{i \frac{(\Om_2 - \Om_1)(U_2 - U_1)}{2}}}{16\pi\sqrt{\Om_1\Om_2} R^2((U_1+U_2)/2, V)} \,
e^{- \frac{(\Delta\Om)^2 (U_2 - U_1)^2}{4}} \,
e^{- \frac{(\Om_1+\Om_2)^2}{4(\Delta\Om)^2}},
\label{49}
\end{align}
where we have assumed that $R(U, V)$ is approximately a constant
in a small neighborhood of the width $\Delta U$
around $U = (U_1+U_2)/2$.
This is a valid assumption as long as $\Delta U \ll a$.

The exponential factor $\exp\left(- (\Delta\Om)^2 (U_2 - U_1)^2/4\right)$ in eq.\eqref{49}
reflects the locality of the interaction.
The last exponential factor $\exp\left(- (\Om_1+\Om_2)^2/4(\Delta\Om)^2\right)$
imposes the requirement of the approximate $P_U$-conservation:
$\Om_1 + \Om_2 \simeq 0$.
The amplitude is exponentially suppressed
when $P_U$-conservation is significantly violated,
i.e. when $(\Om_1 + \Om_2)/\Delta \Om \gg 1$.
As we can choose wave packets with small $\Delta \Om$,
we see that only the states with low frequencies ($\Om \sim 0$)
have a significant probability to be created from the vacuum.
This is a more precise formulation of $P_U$-conservation in quantum theory.

However,
the situation is quite different
from the perspective of distant observers.
In quantum mechanics,
we define the momentum operators
\be
P_u \equiv - i \hbar \del_u,
\quad
P_U \equiv - i \hbar \del_U
\quad \Rightarrow \quad
P_u = \frac{dU}{du} P_U = e^{- \frac{u}{2a}} P_U,
\ee
where we have used eq.\eqref{dUdu}.
The operator $P_u$ does not commute with $P_U$.
Instead,
\be
[P_u, P_U] 
= \frac{i\hbar}{2a} P_U,
%= \frac{i\hbar}{2a} e^{\frac{u}{2a}} P_u,
\ee
which implies
%through eq.\eqref{dom-omp}
the uncertainty relation:
%%% p3
\begin{align}
(\Delta \om) (\Delta \Om) \gtrsim
%\frac{\om}{a} \, e^{\frac{u}{2a}} = 
\frac{\Om}{a},
\label{UR}
\end{align}
%for $P_u = \hbar\om$ and $P_U = \hbar\Om$.
where $\Om$ ($\om$) is the expectation value of $P_U$ ($P_u$) for a given state,
and $\Delta \Om$ ($\Delta \om$) the standard deviation.

From the viewpoint of a distant observer,
the validity of the effective theory on its prediction of Hawking radiation
demands
the existence of the states with wave packets satisfying $\Delta\om \lesssim \om \sim 1/a$.
For such wave-packet states,
we have
\begin{align}
\Delta \Om 
\gtrsim \Om
\label{414}
\end{align}
as a result of eq.\eqref{UR}.
From eq.\eqref{49},
we have seen that $P_U$-conservation only forbids
the creation of particles with
$\Om_1 + \Om_2 + \cdots \gg \Delta \Om$.
Because of eq.\eqref{414},
we conclude that
$P_U$-conservation has little constraint
on outgoing particles with such wave packets $\om \sim \mathcal{O}(1/a) \gtrsim \Delta \om$.
This explains how the detection of particles at distances
is compatible with momentum conservation in the freely falling frame.

Note that the detection of such particles for distant observers
does not imply the detection of high-frequency particles
(those with large $\Om$ and $\Om \gg \Delta \Om$)
for freely falling observers,
\footnote{
This is because a particle with large $\Om$ but $\Delta\Om > \Om$
can have a frequency anywhere in the range $(\Om-\Delta\Om, \Om+\Delta\Om)$,
which can be very small.
}
even though the former is simply a superposition of the latter (like eq.\eqref{2+0}).
That is,
the $S$-matrix is totally different for different wave packets ---
it depends on whether they are suitable for the coordinate $u$ or $U$.
As the creation of high-frequency particles from the vacuum is prohibited by $P_U$-conservation,
the interpretation of the large amplitude of particle creation for distant observers
as a signal of firewall at the horizon in Refs.\cite{Ho:2020cbf,Ho:2020cvn}
is not justified (at least in the analysis so far).
%Whether there will be a firewall
%after the effective theory breaks down,
%on the other hand,
%is an independent question (see also Sec.\ref{Conclusion}).

We are now ready to comment on the nice-slice argument \cite{Lowe:1995ac,Polchinski:1995ta}.
It states that the Hamiltonian evolution of nice time slices 
should not create excitations of energies $\gg \mathcal{O}(1/a)$
according to the adiabatic theorem.
Naively, it implies that
no UV theory is needed here.

As far as we have checked,
the horizon remains uneventful for freely falling observers
as we explained below eq.\eqref{49},
so the effective theory continues to be valid
to describe low-energy physics around the horizon.
Furthermore,
the outgoing particles are of small energies for distant observers.
The nice-slice argument is valid in the sense that
there is no high-energy (real) event on nice slices.

On the other hand,
the large energy scale $E$ \eqref{E>} responsible for large amplitudes
for higher-derivative interactions is locally Lorentz-invariant,
completely independent of the choice of time slices.
If the time slices are chosen such that
the momentum $P_V$ of the infalling matter is small,
the momentum $P_U$ of an outgoing virtual Hawking particle must be large.
(That is the case for the conventional construction of nice slices \cite{Lowe:1995ac}.)
Although this trans-Planckian interaction does not lead to high-energy events
because it is associated with virtual particles,
its proper description still demands a UV theory.

In short,
the nice-slice argument may exclude high-energy (real) events,
but we still need a UV theory to describe processes
involving virtual particles with an invariant trans-Planckian energy.
%even when they involve virtual particles.
In the following,
we shall identify such physical processes,
and show how they are in need of a UV description.

%%%%%%%%%%%%%%%%%%%%%%%%%%%%%%%%%%%%%%%%%%%%%%%%%%
%%%%%%%%%%%%%%%%%%%%%%%%%%%%%%%%%%%%%%%%%%%%%%%%%%
%%%%%%%%%%%%%%%%%%%%%%%%%%%%%%%%%%%%%%%%%%%%%%%%%%

\section{Particle creation by higher-derivative interactions}
\label{HDPC}

In this section,
as an explicit demonstration of the general arguments in Secs.\ref{TPP} and \ref{FFF},
we compute the amplitude of particle creation
due to higher-derivative couplings to the background curvature.
As examples,
we consider
\begin{align}
\hat{\cal O}_n \equiv
\frac{g_n}{2} \,
g^{\mu_1\lam_1} \cdots g^{\mu_n\lam_n} 
g^{\nu_1\rho_1} \cdots g^{\nu_n\rho_n}
R_{\mu_1\nu_1} \cdots R_{\mu_n\nu_n}
(\nabla_{\lam_1} \cdots \nabla_{\lam_n} \phi)
(\nabla_{\rho_1} \cdots \nabla_{\rho_n} \phi),
\label{On-def}
\end{align}
where $g_n$ is the coupling constant
and $R_{\mu\nu}$ the Ricci tensor.
%[The normal ordering is not needed for the calculation below so it is omitted.]
%}
%and $:A:$ the normal ordering of $A$.
%\footnote{
%For the problem at hand,
%it does not matter whether we normal-order with respect to
%$(a_{\Om}, a^{\dag}_{\Om})$ or  $(b_{\om}, b^{\dag}_{\om})$.
The dimension of $g_n$ is $- (4n - 2)$,
so we expect
\begin{align}
g_n \sim \mathcal{O}(1/M_p^{4n-2}).
\label{lam-order}
\end{align}

The background due to the collapsing null shell
is described in Sec.\ref{background}.
The Ricci tensor is determined by the Einstein equation through eq.\eqref{TVV}:
\begin{eqnarray}
%&
%g^{UV} = - 2,
%\quad
%g^{UU} = g^{VV} = 0,
%\nn \\
%&
R_{VV}(U, V) = \frac{a}{R^2} \d_{d}(V - V_s).
%\quad
%R_{UU} = R_{UV} = 0.
\end{eqnarray}
Here,
$\d_{d}(V)$ denotes a regularized $\d$-function.
It has a support of the size $d$ and a peak of the height $\sim 1/d$.
We shall assume that the thickness of the shell $d \ll a$,
%%% p1
and that the characteristic length scales of functions $f(V)$
to be considered below are much larger than $d$,
%%% 1p
so that
\begin{align}
\d_{d}(V - V_s) f(V) &\simeq \d_{d}(V - V_s) f(V_s).
\end{align}
%for any smooth function $f(V)$ of the characteristic length scale $a$.
We also have
\begin{align}
\d_{d}(V - V_s) \, \d_{d}(V - V_s) &\simeq \frac{1}{d} \, \d_{d}(V - V_s).
\end{align}
The interaction \eqref{On-def} can now be approximated by
\begin{align}
\hat{\cal O}_n(U, V) \simeq
\frac{- g_n (-2)^{n-1} a^n}{d^{n-1} R^{2n}(U, V_s)} \,
\d_{d}(V - V_s)
[\nabla_U^n \phi(U, V_s)]^2 \, .
\label{On-1}
\end{align}

The goal of this section is to compute the amplitude
for the pair creation of particles from the Unruh vacuum
for the interaction \eqref{On-def}:
\begin{align}
{\cal M}_n \equiv
\int d^4 x \, \sqrt{-g} \,
\langle f | \hat{\cal O}_n | i \rangle.
\label{amp-pair}
\end{align}
The initial state $| i \rangle$ is the Unruh vacuum $| 0 \rangle$,
and the final state
\begin{align}
| f \rangle \equiv
b^{\dag}_{(\om_1,u_1)}b^{\dag}_{(\om_2,u_2)} | 0 \rangle
\label{f-def}
\end{align}
is a state with 2 extra particles on top of Hawking radiation
for distant observers,
which is equivalently a superposition of 2-particle states in a local freely falling frame
(see eq.\eqref{2+0}).

\detail{
\vtwo{
As physical wave packets should be well localized,
\footnote{
\vtwo{
In general, 
a wave packet well localized in the $r$-space
includes both outgoing and ingoing modes.
The ingoing modes at large distances can be neglected
since they are originated from regions farther away from the black hole
and have no relevance to the black-hole physics.
}
}
we should be allowed to introduce an infra-red cutoff $u_{\infty}$
in the spacetime integral $\int d^4 x$ in ${\cal M}_n$ \eqref{amp-pair},
assuming that $u_{\infty} - u_i \gg \Delta u$ for both $i = 1, 2$.
}
}

%%% p1
To compute the amplitude \eqref{amp-pair},
we need to evaluate
\be
\langle 0 | (\nabla_U^n \phi)^2 b^{\dag}_{(\om_1, u_1)} b^{\dag}_{(\om_2, u_2)} | 0 \rangle.
\ee
Using eqs.\eqref{bdag-wp}, \eqref{2+0} and $a_{\Om} | 0 \rangle = 0$,
it reduces to the product of two factors each of the form
$\langle 0 | \nabla_U^n \phi(U, V) \, b^{\dag}_{(\om_0, u_0)} | 0 \rangle$.
%%% 1p

%%% p1
To estimate $\langle 0 | \nabla_U^n \phi(U, V) \, b^{\dag}_{(\om_0, u_0)} | 0 \rangle$,
%%% 1p
we assume that
\begin{align}
\frac{1}{R(U, V_s)} \ll \Om,
\label{OmR>1}
\end{align}
so that the covariant derivative $\nabla_U$ can be approximated by the partial derivative $\del_U$.
Using eqs.\eqref{U-def}, \eqref{phi-modes}, \eqref{a-CR}, \eqref{alpha}, \eqref{bdag-wp},
as well as the identity
\begin{align}
\int_0^{\infty} \frac{d\Om}{\sqrt{\Om}} \, (i\Om)^n \, \a_{\om\Om} \, e^{i \Om U}
= \frac{A_n(\om)}{(2a)^n} \, \frac{e^{i\om u}}{\sqrt{\om}} \,
\frac{1}{1 - e^{-4\pi a\om}} \, e^{\frac{nu}{2a}}
\qquad (n \geq 1),
\end{align}
where
\begin{align}
A_n(\om) \equiv (n - 1 + 2ia\om)(n - 2 + 2ia\om) \cdots (2ia\om),
\label{An-def}
\end{align}
we find
\begin{align}
\langle 0 | \nabla_U^n \phi(u, v) \, b^{\dag}_{(\om_0, u_0)} | 0 \rangle
&\simeq
\int_0^{\infty} d\om \, f_{\om_0}(\om) \,
\frac{A^{\ast}_n(\om)}{(2a)^n} \, \frac{e^{- i \om (u - u_0)}}{4\pi \sqrt{\om} R} \,
\frac{1}{1 - e^{-4\pi a\om}} \, e^{\frac{nu}{2a}}
\label{bB0}
\\
&=
\frac{1}{4\pi R} \, \frac{1}{(2a)^n} \, 
F^{\ast}_{(n, \om_0)}(u - u_0) \,
e^{\frac{nu}{2a}},
\label{0dnUphib0}
\end{align}
where
\begin{align}
F_{(n, \om_0)}(u) &\equiv
\int_{0}^{\infty} \frac{d\om}{\sqrt{\om}} \,
\frac{f^{\ast}_{\om_0}(\om) A_n(\om)}{1 - e^{-4\pi a\om}} \,
e^{i \om u}.
\label{F-def}
\end{align}

A physical particle should be described by
a wave packet with a finite span in spacetime.
For simplicity, 
we consider a Gaussian wave packet \eqref{f-Gaussian},
for which $F_{(n, \om_0)}(u)$ \eqref{F-def} is approximately a Gaussian
\begin{align}
F_{(n, \om_0)}(u) 
&\simeq
c_{\om_0} \frac{\sqrt{2\pi} (\Delta\om)}{2 \sqrt{\om_0}}
\frac{A_n(\om_0)}{1 - e^{-4\pi a\om_0}} \,
e^{i \om_0 u} \, e^{- \frac{(\Delta\om)^2 u^2}{2}},
\label{516}
\end{align}
assuming that $\om_0 \gg \Delta\om$.
\footnote{
More precisely,
the condition is
$\frac{\om_0}{\Delta\om} \gtrsim \frac{u}{\Delta u} = u\Delta\om$.
}
%The effect of the last term $\exp(nu/2a)$ in eq.\eqref{0dnUphib0}
%is to shift the center of the Gaussian from $u_0$ to $u_0 + n/(2a(\Delta\om)^2)$.

We shall express the amplitude \eqref{amp-pair}
in terms of the central coordinate
\begin{align}
u_c \equiv \frac{u_1 + u_2}{2},
\label{uc-def}
\end{align}
of the two particles,
and the separation
\begin{align}
\d u \equiv u_2 - u_1
\label{du-def}
\end{align}
between them.
We expect that the amplitude is larger when $\d u \simeq 0$
since the two particles are created from a local interaction,
as in eq.\eqref{49}.
We want to understand how the amplitude depends on $u_c$,
e.g., whether the amplitude is larger at a later time.

Using eqs.\eqref{On-1}, \eqref{0dnUphib0}, and \eqref{F-def},
we calculate the amplitude \eqref{amp-pair} as
\begin{align}
%\int d^4 x \, \langle f | {\cal O}_n | i \rangle
{\cal M}_n
&\simeq
\frac{g_n (-1)^n}{4\pi d^{n-1} (2a)^{n}} \,
G_{(n, \om_1, \om_2)}(\d u) \,
e^{\frac{(2n - 1)u_c}{2a}},
\label{ampl-G}
\end{align}
where
\footnote{
It can be shown that the $u$-integral in $G_{(n, \om_1, \om_2)}$ \eqref{G-def} diverges
due to the exponential factor $e^{\frac{(2n - 1) (u - u_c)}{2a}}$.
%in the limit $u_{\infty} \rightarrow \infty$.
Physically,
any detected wave packet is always well localized,
so one is justified to impose an IR cutoff $u_{\infty}$ on the $u$-integral $\int_{-\infty}^{u_{\infty}}$.
For the sake of the simplicity of arguments,
we shall proceed and reach the same conclusion
without worrying about the cutoff.
}
\begin{align}
G_{(n, \om_1, \om_2)}(\d u) &\equiv
\int_{- \infty}^{\infty} du \, \frac{e^{\frac{(2n - 1) (u - u_c)}{2a}}}{R^{2n}(U(u), V_s)} \,
F_{(n, \om_1)}(u - u_c + \d u/2)
F_{(n, \om_2)}(u - u_c - \d u/2).
\label{G-def}
\end{align}
We have used eq.\eqref{U-def} to rewrite the integral of $U$ as an integral of $u$.
\detail{
\vtwo{
and $u_{\infty}$ is the cutoff of the $u$-integral mentioned above.
}
}
%%% p2
At large $u_c$,
the radius $R \simeq a$ is approximately a constant,
and we can shift the integration variable $u$ to absorb $u_c$ in eq.\eqref{G-def},
so that $G_{(n, \om_1, \om_2)}(\d u)$ becomes insensitive to $u_c$.
%%% 2p
Hence,
the amplitude \eqref{ampl-G} increases exponentially with $u_c$ as
\be
{\cal M}_n \propto e^{\frac{(2n - 1)u_c}{2a}}.
\label{ampl}
\ee

Notice that the exponential growth with $u_c$ in the amplitude \eqref{ampl}
is independent of the details of the wave packets,
and that it is a faster exponential growth for a higher-derivative interaction \eqref{On-def} with a larger $n$.
\footnote{
As the amplitude is an invariant quantity,
it is independent of the choice of the coordinate system.
In terms of the asymptotic coordinates $(u, v)$,
the interaction term \eqref{On-def} involves the factor
$(g^{uv})^n \propto (1 - a/r)^{-n}$,
which blows up at the horizon.
In terms of the Kruskal coordinates $(U, V)$,
the derivatives $\del_U^n$ give a factor of $\Om^n$,
where $\Om$ \eqref{dom-omp} grows exponentially
due to the blue-shift factor $(dU/du)^{-1}$.
}
Only the overall factor $G_{(n, \om_1, \om_2)}(\d u)$ in eq.\eqref{ampl-G} depends
on the shape of wave packets.
%\footnote{
%The integral \eqref{F-def} is ill-defined for
%the wave packet Hawking used
%(See footnote \ref{HWP}),
%since it has a tail that goes to $0$ too slowly in the limit $u \rightarrow \infty$.
%}

%As an example,
%for the Gaussian wave packet \eqref{f-Gaussian},
%we can compute the quantity $G_{(n, \om_1, \om_2)}(\d u)$.
Using eq.\eqref{516},
we find
%the magnitude of $G_{(n, \om_1, \om_2)}$ \eqref{G-def} can be estimated as
\begin{align}
%\vtwo{
\left| G_{(n, \om_1, \om_2)}(\d u) \right|
%&
\gtrsim
\frac{\pi^{3/2} \, \left| A_n(\om_1)A_n(\om_2) \right|
%\,e^{i \frac{(2n-1)(\om_1 + \om_2)}{4 a(\Delta\om)^2}}
%e^{\frac{i}{2}(\om_1 - \om_2) \d u}
}
{\sqrt{\om_1 \om_2} \, a^{2n} \left(1 - e^{-4\pi a\om_1}\right)\left(1 - e^{-4\pi a\om_2}\right)} \,
e^{\frac{\left(\frac{2n-1}{2a}\right)^2 - (\om_1 + \om_2)^2}{4(\Delta\om)^2}} \,
e^{- \frac{(\Delta\om)^2 (\d u)^2}{4}}
%}
\label{G-value}
\end{align}
near $R \simeq a$.
The magnitude of the amplitude is then 
determined by eqs.\eqref{ampl-G} and \eqref{G-value} as
\begin{align}
%\left| \int d^4 x \, \langle f | {\cal O}_n | i \rangle\right|
\left|{\cal M}_n\right|
&\gtrsim
\frac{\sqrt{\pi} \, g_n \, \left|A_n(\om_1)A_n(\om_2)\right|}
{d^{n-1} 2^{n+2} \sqrt{\om_1 \om_2} \, a^{3n} \left(1 - e^{-4\pi a\om_1}\right)\left(1 - e^{-4\pi a\om_2}\right)}
\times
\nn \\
&\qquad\qquad \times
e^{\frac{\left(\frac{2n-1}{2a}\right)^2 - (\om_1 + \om_2)^2}{4(\Delta\om)^2}} \,
e^{- \frac{(\Delta\om)^2 (\d u)^2}{4}}
e^{\frac{(2n - 1)u_c}{2a}}.
\label{ampl-pair-2}
\end{align}

Since eq.\eqref{ampl-pair-2} is a result of the low-energy effective theory,
it may not be valid for very high-frequencies $\om_1$, $\om_2$.
However,
whenever Hawking radiation is considered a valid prediction of the effective theory,
states including outgoing particles for distant observers with
\be
\om_1, \om_2 \sim \mathcal{O}(1/a)
\label{om1om2-order}
\ee
must be allowed.
Assuming eq.\eqref{om1om2-order},
the magnitude of the amplitude \eqref{ampl-pair-2} 
satisfies the bound
\footnote{
If eq.\eqref{om1om2-order} is replaced by
$\om_1, \om_2 \gg \mathcal{O}(1/a)$,
the amplitude is larger by a factor of $(a\om_1)^n(a\om_2)^n$.
For the case of
$\om_1, \om_2 \ll \mathcal{O}(1/a)$,
it is smaller by a factor of $(a\om_1)(a\om_2)$.
The discussion below will be essentially the same.
}
\begin{align}
%\left| \int d^4 x \, \langle f | {\cal O}_n | i \rangle\right|%
\left|{\cal M}_n\right|
&\gg
\left(\frac{\ell_p}{a}\right)^{4n-2} \,
e^{\frac{\left(\frac{2n-1}{2a}\right)^2 - (\om_1 + \om_2)^2}{4(\Delta\om)^2}} \,
e^{- \frac{(\Delta\om)^2 (\d u)^2}{4}}
e^{\frac{(2n - 1)u_c}{2a}},
\label{ampl-1-final}
\end{align}
where we have used $d \ll a$.

The factor
$\exp\left\{\left[\left(\frac{2n-1}{2a}\right)^2 - (\om_1 + \om_2)^2\right]/4(\Delta\om)^2\right\}$
in eq.\eqref{ampl-1-final}
%%% p1
is composed of two parts.
The first part $\exp\left[\left(\frac{2n-1}{2a}\right)^2/4(\Delta\om)^2\right]$
is an enhancement due to the higher derivatives in the interaction $\hat{\cal O}_n$ \eqref{On-def}.
The second part $\exp\left[- (\om_1 + \om_2)^2/4(\Delta\om)^2\right]$
is the suppression due to the non-conservation of the $P_u$-momentum.
The joint effect can be of $\mathcal{O}(1)$ or larger
%%% 1p
by choosing $n$ to be sufficiently large.
For instance, 
if $\om_1 = \om_2 = 1/a$,
this factor is larger than $1$ as long as $n \geq 3$.

The next factor $\exp\left(- (\Delta\om)^2 (\d u)^2/4\right)$
in eq.\eqref{ampl-1-final} is expected
as the two particles are created in pairs by a local operator.
The last factor $\exp[(2n-1)u_c/2a]$ grows exponential with 
the central coordinate $u_c$ of the created particles.
Therefore, 
the amplitude for the creation of outgoing particles at large distances
through higher-derivative couplings to the Ricci tensor
grows exponentially with time,
as we have seen in eq.\eqref{ampl-exp}.

The calculation above relies on the Bogoliubov transformation
which suffers the trans-Planckian problem.
Yet,
as long as Hawking radiation is still assumed to be a valid prediction of the effective theory,
the Bogoliubov transformation has to be valid for 
the dominant modes of Hawking radiation, i.e. $\om \sim 1/a$,
and the result \eqref{ampl-1-final} for the case \eqref{om1om2-order} should be valid.
For this case,
we have (as expected above in eq.\eqref{ampl-exp})
\begin{align}
%\left| \int d^4 x \, \langle f | {\cal O}_n | i \rangle\right|
\left| {\cal M}_n \right|
&\gtrsim 
\left[\frac{\ell_p^2}{a^2} \, e^{\frac{u_c}{2a}}\right]^{2n-1}
%= \left(\frac{E^2(u_c)}{M_p^2}\right)^{2n-1},
\label{ampl-0}
\end{align}
for $(2n-1) \geq 2a(\om_1+\om_2)$ and $u_2 \simeq u_1$.
When the magnitude of the transition amplitude is $\sim \mathcal{O}(1)$,
the low-energy effective theory breaks down.
This happens when
\be
u_c \gtrsim u_{B} \equiv 2a \log(a^2/\ell_p^2),
\label{uc-bound}
\ee
which is of the same order as the scrambling time \cite{Sekino:2008he}.
%%% p2
According to eqs.\eqref{U-def}, \eqref{R}, and \eqref{Vs},
the $u$-coordinate is defined such that
the collapsing shell has the radius $R \simeq 2a$ at $u = 0$.
Thus, its radius satisfies
\be
R(u_B, v_s) - a \simeq \ell_p^2/a
\ee
at the scrambling time.
%%% 2p
%%% p3
\hide{
The blue-shift factor at the scrambling time is
\be
\left(\frac{dU(u_B)}{du}\right)^{-1} = \frac{a^2}{\ell_p^2}.
\label{dUdu-us}
\ee
}
%%% 3p

%%% p3
We conclude that,
by the scrambling time $u_B$
after passing through the point $R = 2a$
(equivalently, 
when the matter shell has $R = a + \mathcal{O}(\ell_p^2/a)$),
the amplitude increases to $\mathcal{O}(1)$,
and the effective theory breaks down
for the outgoing modes with $\om \gtrsim \mathcal{O}(1/a)$.
It is thus also the last moment
when Hawking radiation is a valid prediction of the effective theory.
%%% 3p

%%%%%%%%%%%%%%%%%%%%%%%%%%%%%%%%%%%%%%%%%%%%%%%%%%
%%%%%%%%%%%%%%%%%%%%%%%%%%%%%%%%%%%%%%%%%%%%%%%%%%
%%%%%%%%%%%%%%%%%%%%%%%%%%%%%%%%%%%%%%%%%%%%%%%%%%

\section{Infalling particle radiation}
\label{TAIP}

In this section,
as another explicit example of the general discussions in Secs.\ref{TPP} and \ref{FFF},
we compute the transition amplitude for the decay of an infalling massless particle
via 4-point higher-derivative interactions.
It includes the case of an infalling particle radiating two outgoing particles.
%\footnote{
%This happens when the particle in the in-state is identified with a particle in the out-state.
%}
The outgoing particles are defined with respect to distant observers
as in Sec.\ref{HDPC}.
\footnote{
We can think of the particle production via couplings to the Ricci tensor in Sec.\ref{HDPC}
as an analogue of the radiation by infalling particles in this section
%Sec.\ref{TAIP}
in the sense that the radiation through direction interaction in the latter
is replaced by an indirect interaction mediated by gravity in the former.
}

Introduce several massless scalar fields
$\phi_A$ and $\psi_A$ ($A = 1, 2$),
each with the same mode expansion as $\phi$ \eqref{phi-modes}.
For our discussion below,
we will only need the ingoing modes of $\phi_A$
and the outgoing modes of $\psi_A$,
and the corresponding creation-annihilation operators are denoted
$(\tilde{a}_{A\Om}, \tilde{a}^{\dag}_{A\Om})$
and
$(b_{A\om}, b^{\dag}_{A\om})$,
respectively.

We consider the following higher-derivative interactions among these fields:
\begin{align}
\hat{\cal O}_{mn} &\equiv
\lam_{mn} g^{\mu_1\nu_1} \cdots g^{\mu_m\nu_m}
g^{\lam_1\rho_1} \cdots g^{\lam_n\rho_n}
: (\nabla_{\mu_1}\cdots\nabla_{\mu_m}\phi_1)
(\nabla_{\lam_1}\cdots\nabla_{\lam_n}\phi_2) \,
\times
\nn \\
&\qquad \times \,
(\nabla_{\nu_1}\cdots\nabla_{\nu_m}\psi_1)
(\nabla_{\rho_1}\cdots\nabla_{\rho_n}\psi_2) :,
\label{Lint}
\end{align}
where
\be
\lam_{mn} \sim \mathcal{O}(1/M_p^{2(m+n)})
\ee
is the coupling constant.
With this class of higher-derivative interactions,
we now compute the transition amplitude
for an ingoing particle $\phi_1$ to decay into
an ingoing particle $\phi_2$ and two outgoing particles $\psi_1$ and $\psi_2$.
The initial and final states are
\detail{
\footnote{
For the ingoing modes,
the difference between particles defined with respect to
the coordinates $V$ and $v$ is not essential,
since the exponential relation \eqref{V-def}
only applies to $V \geq V_s = 2a$,
where the blue-shift factor is never very large.
%%% p3
The result should be essentially the same
if we replace $\tilde{a}_{A\Om}$ by $\tilde{b}_{A\om}$
in the initial and final states.
%%% 3p
}
}
\begin{align}
| i \rangle &= \tilde{a}^{\dag}_{1\Om} | 0 \rangle,
\label{i2}
\\
| f \rangle &= b^{\dag}_{1(\om_1, u_1)} b^{\dag}_{2(\om_2, u_2)}
\tilde{a}^{\dag}_{2\Om'} | 0 \rangle,
\label{f2}
\end{align}
%%% p1
and we shall calculate the amplitude
\be
{\cal M}_{mn} \equiv
\int d^4 x \, \sqrt{-g} \, \langle f | \hat{\cal O}_{mn} | i \rangle.
\label{65}
\ee
%%% 1p

For this process,
we can simplify the interaction \eqref{Lint} as
\be
\hat{\cal O}_{mn}
\simeq
\lam_{mn}
(g^{UV})^{m+n} (\del_V^m \phi_1)(\del_V^n \phi_2)(\del_U^m \psi_1)(\del_U^n \psi_2),
\ee
assuming that eq.\eqref{OmR>1} applies to all the relevant frequencies.
Then,
we just have to calculate
\be
\langle f | \hat{\cal O}_{mn} | i \rangle \simeq
\lam_{mn} (g^{UV})^{m+n}
\langle 0 | \del_V^m\phi_1 \tilde{a}^{\dag}_{1\Om} | 0 \rangle
\langle 0 | \del_V^n\phi_2 \tilde{a}^{\dag}_{2\Om'} | 0 \rangle^{\ast}
\langle 0 | \del_U^m\psi_1 b^{\dag}_{1(\om_1,u_1)} | 0 \rangle^{\ast}
\langle 0 | \del_U^n\psi_2 b^{\dag}_{2(\om_2,u_2)} | 0 \rangle^{\ast}.
\ee

%Analogous to eq.\eqref{0dnUphib0},
For the ingoing modes, 
we have
\begin{align}
\langle 0 | \del_V^n \phi_A(U, V) \tilde{a}^{\dag}_{B\Om} | 0 \rangle
&=
\d_{AB} \,
\frac{(- i\Om)^n}{4\pi\sqrt{\Om} R} \, e^{- i \Om V}.
\label{aA0}
\end{align}
(For the ingoing modes,
it makes no significant difference whether the wave packets are introduced,
as long as their frequencies $\Om$
have small uncertainties $\Delta\Om \ll \Om$.)
Together,
eqs.\eqref{0dnUphib0} and \eqref{aA0}
allow us to express the amplitude as
\begin{align}
&
{\cal M}_{mn}
= i^{m-n} \lam_{mn} \, \frac{\Om^{m+n-1} \d(\Om - \Om')}{32 \pi^2 a^{m+n+2}} \, 
e^{(m+n-1)\frac{u_c}{2a}} \, H_{(m, n,\om_1,\om_2)}(\d u),
\end{align}
where we have assumed $R \simeq a$ and used eqs.\eqref{F-def}, \eqref{uc-def}, \eqref{du-def},
and
\be
H_{(m, n,\om_1,\om_2)}(\d u) \equiv \int_{-\infty}^{\infty} du \,
e^{(m+n-1)\frac{u}{2a}} \,
F_{(m,\om_1)}(u + \d u/2) F_{(n,\om_2)}(u - \d u/2).
\ee

Similar to $G_{(n, \om_1, \om_2)}$ \eqref{G-value},
$H_{(m, n, \om_1, \om_2)}(\d u)$ can be evaluated
for the Gaussian wave packets \eqref{f-Gaussian} as
\begin{align}
|{\cal M}_{mn}|
&\sim
(a\Om)^{m+n-1} 
\left(\frac{\ell_p^2}{a^2}\right)^{m+n} \d(\Om - \Om') \,
e^{\frac{\left(\frac{m+n-1}{2a}\right)^2 - (\om_1 + \om_2)^2}{4(\Delta\om)^2}} \,
e^{- \frac{(\Delta\om)^2 (u_2 - u_1)^2}{4}} \, 
e^{(m+n-1)\frac{u_c}{2a}},
\label{trans-0}
\end{align}
where as in Sec.\ref{HDPC}
we have assumed
\detail{
\footnote{
We consider these frequencies because they should be present
in the low-energy effective theory whenever
its prediction of Hawking radiation is still valid.
}
}
%%% p3
\be
\om_1 \sim \om_2 \sim \mathcal{O}(1/a).
\ee
For the particle $\phi_1$ to fall into the black hole,
its wavelength is restricted to be smaller than $a$,
so we have $\Om \gtrsim 1/a$,
and eq.\eqref{trans-0} gives a bound on the amplitude as
\begin{align}
|{\cal M}_{mn}|
&\gtrsim
\left(\frac{\ell_p^2}{a^2}\right)^{m+n} \d(\Om - \Om') \,
e^{\frac{\left(\frac{m+n-1}{2a}\right)^2 - (\om_1 + \om_2)^2}{4(\Delta\om)^2}} \,
e^{- \frac{(\Delta\om)^2 (u_2 - u_1)^2}{4}} \, 
e^{(m+n-1)\frac{u_c}{2a}}.
\label{trans-1}
\end{align}
%%% 3p

The interpretation of the last 3 factors in eq.\eqref{trans-1}
is analogous to those in eq.\eqref{ampl-1-final}.
In short,
there is an exponentially increasing amplitude
for the infalling particle to send a signal to distant observers
about its presence in the form of two coincident particles $\psi_1$ and $\psi_2$,
%%% p3
and the amplitude becomes of $\mathcal{O}(1)$ at the scrambling time.
%%% 3p

The Dirac $\d$-function in eq.\eqref{trans-1} demands that
the particle $\phi_1$ passes all of its momentum to the particle $\phi_2$,
as a result of the translation symmetry in $V$.
When $\phi_2 = \phi_1$,
this is the amplitude for an infalling particle to radiate two outgoing particles $\psi_1, \psi_2$.
%The infalling particle stays at rest in its comoving freely falling frame when it radiates.
%%% p2
The radiation has no back-reaction,
analogous to the radiation of a freely falling point charge \cite{NoReaction}.
%This intriguing result is analogous to
%the classical manifestation of the equivalence principle
%that a freely falling point charge radiates without back-reaction \cite{NoReaction}.
%We will comment more on this in Sec.\ref{UE}.
%%% 2p

%%% p2
\hide{
%This question has a standard answer for electromagnetic radiation \cite{NoReaction}.
The electromagnetic field of a charge
can be decomposed into the radiation field and the Coulomb field.
The Coulomb field depends on the velocity of the particle,
and the radiation field is present only when the acceleration is non-zero.
From the viewpoint of its comoving freely falling frame,
a point charge in free fall has a constant Coulomb field.
If we think of the accelerated frame as a continuous transition
from an inertial frame to another inertial frame,
the Coulomb field changes under Lorentz boosts,
and the change in its energy accounts for the energy transferred into radiation.
The same interpretation can apply to other interactions as well.
}
%%% 2p

The higher-derivative interaction thus provides a mechanism to transfer
(at least part of) the information of the collapsing matter into out-going particles.
%without violation of the equivalence principle.
The time it takes for the the information transfer is roughly speaking the scrambling time.
This is reminiscent of the original meaning of the scrambling time \cite{Sekino:2008he}.

How much information of the ingoing particle can be transferred to large distances?
If the theory includes both interactions
$\del_V^m \phi_1\del_V^n \phi_1 \del_U^m \psi_1 \del_U^n \psi_2$
and 
$\del_V^m \phi_2\del_V^n \phi_2 \del_U^m \psi_1 \del_U^n \psi_2$,
when coincident outgoing particles $\psi_1$ and $\psi_2$ are detected,
the infalling particle could be either $\phi_1$ or $\phi_2$.
It is interesting to contemplate the possibility that
all information of the infalling particles can be retrieved this way
for certain UV theories.

%%%%%%%%%%%%%%%%%%%%%%%%%%%%%%%%%%%%%%%%%%%%%%%%%%
%%%%%%%%%%%%%%%%%%%%%%%%%%%%%%%%%%%%%%%%%%%%%%%%%%
%%%%%%%%%%%%%%%%%%%%%%%%%%%%%%%%%%%%%%%%%%%%%%%%%%

%%%%%%%%%%%%%%%%%%%%%%%%%%%%%%%%%%%%%%%%%%%%%%%%%%
%%%%%%%%%%%%%%%%%%%%%%%%%%%%%%%%%%%%%%%%%%%%%%%%%%
%%%%%%%%%%%%%%%%%%%%%%%%%%%%%%%%%%%%%%%%%%%%%%%%%%

%%%%%%%%%%%%%%%%%%%%%%%%%%%%%%%%%%%%%%%%%%%%%%%%%%
%%%%%%%%%%%%%%%%%%%%%%%%%%%%%%%%%%%%%%%%%%%%%%%%%%
%%%%%%%%%%%%%%%%%%%%%%%%%%%%%%%%%%%%%%%%%%%%%%%%%%

\section{Conclusion and discussion}
\label{Conclusion}

%\subsection{Generalizations}

In the above,
we have considered two specific types of higher-derivative interactions \eqref{On-def} and \eqref{Lint}.
The conclusion about exponentially increasing amplitudes
(like eq.\eqref{ampl-exp} with $m > 0$)
applies to a much wider class of higher-derivative interactions.
For example,
we can consider more general higher-derivative interactions built from
the following invariant (dimensionless) building blocks:
\begin{align}
&M_p^4 \int d^4 x \sqrt{-g} = M_p^4 \int dU dV 4\pi R^2
\rightarrow
\frac{a^4}{\ell_p^4} \, e^{- \frac{u}{2a}},
\\
&\frac{1}{M_p^2} \, g^{UV} \del_U \otimes \del_V
\rightarrow
e^{\frac{u}{2a}} \frac{\ell_p^2}{a^2},
\\
&\frac{1}{M_p^4} R_{VV}g^{UV}g^{UV} \del_U \otimes \del_U
\rightarrow
\left(\frac{\ell_p^2}{a^2} \, e^{\frac{u}{2a}}\right)^2,
\\
&\frac{1}{M_p} \phi
\rightarrow \frac{\ell_p}{a},
\end{align}
where we have used
$\del_u \sim \del_v \rightarrow 1/a$, $R \rightarrow a$,
$dU/du \rightarrow e^{- u/2a}$, and $dV/dv \rightarrow 1$
to estimate their contributions to the amplitudes.
We can use the formulas above for a rough order-of-magnitude estimate
of the amplitudes.

For a generic combination of these factors,
an amplitude is 
\begin{align}
&\lam \int d^4 x \, \sqrt{-g}
\left[g^{UV} \del_U \otimes \del_V\right]^m
\left[R_{VV} g^{UV} g^{UV} \del_U \otimes \del_U\right]^n
\left[\phi\right]^{\otimes q}
\; \sim \;
\frac{\ell_p^{q-2}}{a^{q-2}}
\left[\frac{\ell_p^2}{a^2} \, e^{\frac{u}{2a}}\right]^{m + 2n - 1},
\end{align}
where the coupling constant $\lam \sim \mathcal{O}(M_p^{4-2m-4n-q})$.
For any finite $m$, $n$, and $q$ as long as $m + 2n > 1$,
it becomes large at the scrambling time.

Note that the onset of the breakdown of the effective theory
is robustly at the scrambling time.
For instance,
by increasing the value of $q$,
the critical time when a given interaction becomes $\mathcal{O}(1)$
is postponed by an amount $\sim \mathcal{O}(2a\log(a/\ell_p))$,
but it is still the same order of magnitude as the scrambling time.

Another type of generalization is to consider other final states.
For instance,
if we replace the 1-particles state $b^{\dag}_{\om}|0\rangle$ by $b_{\om}|0\rangle$,
the wave function \eqref{bB0} is replaced by
\begin{align}
\langle 0 | \del_U^n \phi \, b_{(\om_0, u_0)} | 0 \rangle
&\simeq - \int_0^{\infty} d\om f_{\om_0}(\om)
\frac{A^{\ast}_n(- \om)}{(2a)^n} \, \frac{e^{i \om (u - u_0)}}{4\pi \sqrt{\om} R} \,
\frac{e^{-4\pi a\om}}{1- e^{- 4\pi a\om}} \, e^{\frac{nu}{2a}}.
\end{align}
The same exponential factor $e^{\frac{nu}{2a}}$ appears here.
%We expect the same exponential growth for generic higher-derivative amplitudes.
%In Sec.\ref{HDPC},
%we have shown that the amplitude for particle creation
%on top of Hawking radiation grows exponentially with time.
It is then straightforward to see that
the same higher-derivative interaction would also
have an exponentially growing amplitude
for ``removing'' particles
(with the operator $b_{(\om_0, u_0)}$) from Hawking radiation.

%%% new
Our conclusion is that the effective theory breaks down for
any physical process involving outgoing particles with $\om \gtrsim 1/a$
(including Hawking radiation)
when the collapsing matter is around a Planck length from the horizon.
This is reminiscent of the brick-wall model \cite{trans-Planckian-1,tHooft:1996rdg}
(although the horizon may still remain uneventful
for freely falling observers in our discussion).
It will be interesting to explore further connection between our results
and the brick-wall model,
including its holographic dual \cite{Kay:2011np,Iizuka:2013kma,Terashima:2021klf}.

Let us now speculate on what could happen after the effective theory breaks down.
%It is up to the UV theory how eventful the initially uneventful horizon may become.
A simple possibility is that the UV theory imposes an effective cutoff
that stops both Hawking radiation and the large amplitude for particle creation.
The black hole becomes eternal,
perhaps as a ``fuzzball'' \cite{FuzzBall}.
It is also possible that an effective theory will be valid
after a transition through a UV-mechanism,
possibly with the Boulware vacuum at the horizon.

Another possibility is that the UV theory
keeps the large (but now regulated) amplitude of particle creation,
and a large outgoing energy flux
(which can be identified with the ``firewall'' \cite{firewall,firewall-B})
appears around the horizon.

In addition,
when an outgoing energy flux appears around the horizon,
the spacetime geometry is modified.
The Schwarzschild geometry is no longer applicable.
Such a scenario was proposed in Refs.\cite{Kawai:2013mda,Kawai:2014afa}.
Radiation from the collapsing matter induces
a large tangential pressure on the collapsing shell \cite{Kawai:2020rmt}.
As alternative approaches,
connections between stress energy tensor and geometry
via the semi-classical Einstein equation have been explored
in Refs.\cite{Ho:2018jkm,Terno:2019kwm,Ho:2019kte,ShortDistance,Murk:2021cla}.

It will be interesting to find the relation between
different scenarios for the black-hole evaporation and
the corresponding features in the UV theory.
We leave this problem for future investigation.

To conclude,
we have established in this paper that
the effective-field-theoretic derivation of Hawking radiation
is no longer reliable after the scrambling time.
Planckian physics decides what happens next. 
The original argument for information loss due to Hawking radiation
based on effective-theory calculations can be dismissed,
so the original paradox is resolved in this sense.
On the other hand,
it is still unclear by what kind of UV mechanism
the information is preserved.
The Planckian information problem persists.

%%%%%%%%%%%%%%%%%%%%%%%%%%%%%%%%%%%%%%%%%%%%%%%%%%
%%%%%%%%%%%%%%%%%%%%%%%%%%%%%%%%%%%%%%%%%%%%%%%%%%
%%%%%%%%%%%%%%%%%%%%%%%%%%%%%%%%%%%%%%%%%%%%%%%%%%

\section*{Acknowledgement}

We thank Heng-Yu Chen, Jiunn-Wei Chen, Yu-tin Huang, Samir Mathur, Nobuyoshi Ohta, Suguru Okumura,
Wei-Hsiang Shao, Hideaki Takabe, and Naoki Watamura for valuable discussions.
P.M.H.\ is supported in part by the Ministry of Science and Technology, R.O.C.
(MOST 110-2112-M-002 -016 -MY3) and by National Taiwan University.
H.K. thanks Professor Shin-Nan Yang and his family
for their kind support through the Chin-Yu chair professorship.
H.K. is also partially supported by Japan Society of Promotion of Science (JSPS),
Grants-in-Aid for Scientific Research (KAKENHI) Grants No.\ 20K03970 and 18H03708,
by the Ministry of Science and Technology, R.O.C. (MOST 110-2811-M-002-500),
and by National Taiwan University. 
Y.Y.\ is partially supported by Japan Society of Promotion of Science (JSPS), 
Grants-in-Aid for Scientific Research (KAKENHI) Grants No.\ 21K13929, 18K13550 and 17H01148. 
Y.Y.\ is also partially supported by RIKEN iTHEMS Program.

% End of detail

%%%%%%%%%%%%%%%%%%%%%%%%%%%%%%%%%%%%%%%%%%%%%%%%%%
%%%%%%%%%%%%%%%%%%%%%%%%%%%%%%%%%%%%%%%%%%%%%%%%%%
%%%%%%%%%%%%%%%%%%%%%%%%%%%%%%%%%%%%%%%%%%%%%%%%%%

%\end{CJK} 

\vskip .8cm
\baselineskip 22pt
\begin{thebibliography}{99}
\itemsep 0pt






%\cite{Mathur:2009hf}
\bibitem{Mathur:2009hf} 
  S.~D.~Mathur,
  ``The Information paradox: A Pedagogical introduction,''
  Class.\ Quant.\ Grav.\  {\bf 26}, 224001 (2009)
  [arXiv:0909.1038 [hep-th]].
  %%CITATION = ARXIV:0909.1038;%%
  %183 citations counted in INSPIRE as of 24 Apr 2015

\bibitem{firewall}
%\cite{Almheiri:2012rt}
%\bibitem{Almheiri:2012rt} 
  A.~Almheiri, D.~Marolf, J.~Polchinski and J.~Sully,
  ``Black Holes: Complementarity or Firewalls?,''
  JHEP {\bf 1302}, 062 (2013)
  [arXiv:1207.3123 [hep-th]];
  %%CITATION = ARXIV:1207.3123;%%
  %314 citations counted in INSPIRE as of 31 Mar 2015

\bibitem{Hawking:1976ra} 
  S.~W.~Hawking,
  ``Breakdown of Predictability in Gravitational Collapse,''
  Phys.\ Rev.\ D {\bf 14}, 2460 (1976).
  %%CITATION = PHRVA,D14,2460;%%
  %1087 citations counted in INSPIRE as of 23 Apr 2015

\detail{
%\cite{Johnson:2018yci}
\bibitem{Johnson:2018yci}
T.~Johnson,
``On the linear stability of the Schwarzschild solution to gravitational perturbations in the generalised wave gauge,''
[arXiv:1803.04012 [gr-qc]].
%13 citations counted in INSPIRE as of 09 Sep 2021

%\cite{Dafermos:2014cua}
\bibitem{Dafermos:2014cua}
M.~Dafermos, I.~Rodnianski and Y.~Shlapentokh-Rothman,
``Decay for solutions of the wave equation on Kerr exterior spacetimes III: The full subextremal case |a| \ensuremath{<} M,''
[arXiv:1402.7034 [gr-qc]].
%80 citations counted in INSPIRE as of 09 Sep 2021

%\cite{Hadar:2017ven}
\bibitem{Hadar:2017ven}
S.~Hadar and H.~S.~Reall,
``Is there a breakdown of effective field theory at the horizon of an extremal black hole?,''
JHEP \textbf{12}, 062 (2017)
doi:10.1007/JHEP12(2017)062
[arXiv:1709.09668 [hep-th]].
%17 citations counted in INSPIRE as of 09 Sep 2021
}

%\cite{Dodelson:2015toa}
\bibitem{Dodelson:2015toa}
M.~Dodelson and E.~Silverstein,
``String-theoretic breakdown of effective field theory near black hole horizons,''
Phys. Rev. D \textbf{96}, no.6, 066010 (2017)
doi:10.1103/PhysRevD.96.066010
[arXiv:1504.05536 [hep-th]].
%29 citations counted in INSPIRE as of 18 Apr 2020

%\cite{Weinberg:1995mt}
\bibitem{Weinberg:1995mt}
S.~Weinberg,
``The Quantum theory of fields. Vol. 1: Foundations,''
%334 citations counted in INSPIRE as of 02 Apr 2020

%\cite{Iso:2008sq}
\bibitem{Iso:2008sq}
S.~Iso,
``Hawking Radiation, Gravitational Anomaly and Conformal Symmetry: The Origin of Universality,''
Int. J. Mod. Phys. A \textbf{23}, 2082-2090 (2008)
doi:10.1142/S0217751X08040627
[arXiv:0804.0652 [hep-th]].
%21 citations counted in INSPIRE as of 07 Dec 2021

\bibitem{trans-Planckian-1}
%\cite{tHooft:1984kcu}
%\bibitem{tHooft:1984kcu} 
  G.~'t Hooft,
  ``On the Quantum Structure of a Black Hole,''
  Nucl.\ Phys.\ B {\bf 256}, 727 (1985).
  doi:10.1016/0550-3213(85)90418-3
  %%CITATION = doi:10.1016/0550-3213(85)90418-3;%%
  %1078 citations counted in INSPIRE as of 15 Jan 2020

\bibitem{trans-Planckian-2}
%\cite{Jacobson:1991gr}
%\bibitem{Jacobson:1991gr} 
  T.~Jacobson,
  ``Black hole evaporation and ultrashort distances,''
  Phys.\ Rev.\ D {\bf 44}, 1731 (1991).
  doi:10.1103/PhysRevD.44.1731
  %%CITATION = doi:10.1103/PhysRevD.44.1731;%%
  %268 citations counted in INSPIRE as of 15 Jan 2020

%%% Hawking-Radiation %%%
%\cite{Hawking:1974rv}
\bibitem{Hawking:1974rv}
S.~W.~Hawking,
``Black hole explosions,''
Nature \textbf{248}, 30-31 (1974)
doi:10.1038/248030a0
%3427 citations counted in INSPIRE as of 02 May 2021

%\cite{Hawking:1974sw}
\bibitem{Hawking:1974sw} 
  S.~W.~Hawking,
  ``Particle Creation by Black Holes,''
  Commun.\ Math.\ Phys.\  {\bf 43}, 199 (1975)
  [Commun.\ Math.\ Phys.\  {\bf 46}, 206 (1976)].
  %%CITATION = CMPHA,43,199;%%
  %5598 citations counted in INSPIRE as of 31 Aug 2015

%%% Hawking radiation and black hole entropy %%%
%\cite{Hawking:1976de}
\bibitem{Hawking:1976de} 
  S.~W.~Hawking,
  ``Black Holes and Thermodynamics,''
  Phys.\ Rev.\ D {\bf 13}, 191 (1976).
  doi:10.1103/PhysRevD.13.191
  %%CITATION = doi:10.1103/PhysRevD.13.191;%%
  %671 citations counted in INSPIRE as of 13 May 2016

%\cite{Lafrance:1994in}
\bibitem{Lafrance:1994in}
R.~Lafrance and R.~C.~Myers,
``Gravity's rainbow,''
Phys. Rev. D \textbf{51}, 2584-2590 (1995)
doi:10.1103/PhysRevD.51.2584
[arXiv:hep-th/9411018 [hep-th]].
%70 citations counted in INSPIRE as of 18 Apr 2020

%\cite{Sekino:2008he}
\bibitem{Sekino:2008he}
Y.~Sekino and L.~Susskind,
``Fast Scramblers,''
JHEP \textbf{10}, 065 (2008)
doi:10.1088/1126-6708/2008/10/065
[arXiv:0808.2096 [hep-th]].
%510 citations counted in INSPIRE as of 18 Apr 2020

%\cite{Brout:1995rd}
\bibitem{Brout:1995rd} 
  R.~Brout, S.~Massar, R.~Parentani and P.~Spindel,
  ``A Primer for black hole quantum physics,''
  Phys.\ Rept.\  {\bf 260}, 329 (1995)
  doi:10.1016/0370-1573(95)00008-5
  [arXiv:0710.4345 [gr-qc]].
  %%CITATION = doi:10.1016/0370-1573(95)00008-5;%%
  %240 citations counted in INSPIRE as of 16 Feb 2020

\bibitem{trans-Planckian-3}
%\cite{Unruh:1994je}
%\bibitem{Unruh:1994je} 
  W.~G.~Unruh,
  ``Sonic analog of black holes and the effects of high frequencies on black hole evaporation,''
  Phys.\ Rev.\ D {\bf 51}, 2827 (1995).
  doi:10.1103/PhysRevD.51.2827
  %%CITATION = doi:10.1103/PhysRevD.51.2827;%%
  %501 citations counted in INSPIRE as of 15 Jan 2020
%\cite{Brout:1995wp}
%\bibitem{Brout:1995wp} 
  R.~Brout, S.~Massar, R.~Parentani and P.~Spindel,
  ``Hawking radiation without transPlanckian frequencies,''
  Phys.\ Rev.\ D {\bf 52}, 4559 (1995)
  doi:10.1103/PhysRevD.52.4559
  [hep-th/9506121].
  %%CITATION = doi:10.1103/PhysRevD.52.4559;%%
  %179 citations counted in INSPIRE as of 15 Jan 2020
  
\bibitem{trans-Planckian-4}
%\cite{Helfer:2003va}
%\bibitem{Helfer:2003va} 
  A.~D.~Helfer,
  ``Do black holes radiate?,''
  Rept.\ Prog.\ Phys.\  {\bf 66}, 943 (2003)
  doi:10.1088/0034-4885/66/6/202
  [gr-qc/0304042].
  %%CITATION = doi:10.1088/0034-4885/66/6/202;%%
  %56 citations counted in INSPIRE as of 15 Jan 2020papers

%\cite{Leahy:1983vb}
\bibitem{Leahy:1983vb}
D.~A.~Leahy and W.~G.~Unruh,
``EFFECTS OF A LAMBDA PHI**4 INTERACTION ON BLACK HOLE EVAPORATION IN TWO-DIMENSIONS,''
Phys. Rev. D \textbf{28}, 694-702 (1983)
doi:10.1103/PhysRevD.28.694
%12 citations counted in INSPIRE as of 31 May 2021

%\cite{Helfer:2005wz}
\bibitem{Helfer:2005wz}
A.~D.~Helfer,
``Quantum character of black holes,''
[arXiv:gr-qc/0503053 [gr-qc]].
%1 citations counted in INSPIRE as of 31 May 2021

%\cite{Frasca:2014gua}
\bibitem{Frasca:2014gua}
M.~Frasca,
``Hawking radiation and interacting fields,''
Eur. Phys. J. Plus \textbf{132}, no.11, 467 (2017)
doi:10.1140/epjp/i2017-11732-1
[arXiv:1412.1955 [gr-qc]].
%1 citations counted in INSPIRE as of 31 May 2021

%\cite{Ho:2020cbf}
\bibitem{Ho:2020cbf}
P.~M.~Ho and Y.~Yokokura,
``Firewall From Effective Field Theory,''
[arXiv:2004.04956 [hep-th]].
%5 citations counted in INSPIRE as of 09 May 2021

%\cite{Ho:2020cvn}
\bibitem{Ho:2020cvn}
P.~M.~Ho,
``From Uneventful Horizon to Firewall in D-Dimensional Effective Theory,''
[arXiv:2005.03817 [hep-th]].
%2 citations counted in INSPIRE as of 09 May 2021


%\cite{Davies:1976ei}
\bibitem{Davies:1976ei}
P.~Davies, S.~Fulling and W.~Unruh,
``Energy Momentum Tensor Near an Evaporating Black Hole,''
Phys.\ Rev.\ D \textbf{13}, 2720-2723 (1976)
doi:10.1103/PhysRevD.13.2720
%271 citations counted in INSPIRE as of 02 Apr 2020

%\cite{Christensen:1977jc}
\bibitem{Christensen:1977jc} 
  S.~M.~Christensen and S.~A.~Fulling,
  ``Trace Anomalies and the Hawking Effect,''
  Phys.\ Rev.\ D {\bf 15}, 2088 (1977).
  doi:10.1103/PhysRevD.15.2088
  %%CITATION = doi:10.1103/PhysRevD.15.2088;%%
  %454 citations counted in INSPIRE as of 16 Dec 2016

%%% Nice Slice Argument %%%
%\cite{Lowe:1995ac}
\bibitem{Lowe:1995ac}
D.~A.~Lowe, J.~Polchinski, L.~Susskind, L.~Thorlacius and J.~Uglum,
``Black hole complementarity versus locality,''
Phys. Rev. D \textbf{52}, 6997-7010 (1995)
doi:10.1103/PhysRevD.52.6997
[arXiv:hep-th/9506138 [hep-th]].
%190 citations counted in INSPIRE as of 09 Dec 2021

%\cite{Polchinski:1995ta}
\bibitem{Polchinski:1995ta}
J.~Polchinski,
``String theory and black hole complementarity,''
[arXiv:hep-th/9507094 [hep-th]].
%49 citations counted in INSPIRE as of 09 Dec 2021

\bibitem{NoReaction}
T.~Fulton, F.~Rohrlich, 
``Classical radiation from a uniformly accelerated charge,''
Annals of Physics \textbf{9}, 499 (1960);
%\cite{DeWitt:1964de}
%\bibitem{DeWitt:1964de}
C.~M.~DeWitt and B.~S.~DeWitt,
``Falling charges,''
Physics Physique Fizika \textbf{1}, 3-20 (1964)
[erratum: Physics Physique Fizika \textbf{1}, 145 (1964)];
%doi:10.1103/PhysicsPhysiqueFizika.1.3;
%115 citations counted in INSPIRE as of 08 Sep 2021
J.~Cohn, Am. J. Phys. \textbf{46}, 225 (1978);
Chap. 8 or R. ~Peierls, {\em Surprises in Theoretical Physics}
(Princeton: Princeton University Press 1979);
and the article by P.~Pearle in
{\em Electromagnetism: Paths to Research},
ed. D.~Teplitz (New York: Plenum Press, 1982).

%\cite{tHooft:1996rdg}
\bibitem{tHooft:1996rdg}
G.~'t Hooft,
``The Scattering matrix approach for the quantum black hole: An Overview,''
Int. J. Mod. Phys. A \textbf{11}, 4623-4688 (1996)
doi:10.1142/S0217751X96002145
[arXiv:gr-qc/9607022 [gr-qc]].
%295 citations counted in INSPIRE as of 14 Nov 2021

%\cite{Kay:2011np}
\bibitem{Kay:2011np}
B.~S.~Kay and L.~Ort\'\i{}z,
``Brick Walls and AdS/CFT,''
Gen. Rel. Grav. \textbf{46}, 1727 (2014)
doi:10.1007/s10714-014-1727-x
[arXiv:1111.6429 [hep-th]].
%15 citations counted in INSPIRE as of 14 Nov 2021

%\cite{Iizuka:2013kma}
\bibitem{Iizuka:2013kma}
N.~Iizuka and S.~Terashima,
``Brick Walls for Black Holes in AdS/CFT,''
Nucl. Phys. B \textbf{895}, 1-32 (2015)
doi:10.1016/j.nuclphysb.2015.03.018
[arXiv:1307.5933 [hep-th]].
%10 citations counted in INSPIRE as of 14 Nov 2021

%\cite{Terashima:2021klf}
\bibitem{Terashima:2021klf}
S.~Terashima,
``Simple Bulk Reconstruction in AdS/CFT Correspondence,''
[arXiv:2104.11743 [hep-th]].
%0 citations counted in INSPIRE as of 14 Nov 2021

\bibitem{FuzzBall}
%\cite{Lunin:2001jy}
%\bibitem{Lunin:2001jy} 
  O.~Lunin and S.~D.~Mathur,
  ``AdS / CFT duality and the black hole information paradox,''
  Nucl.\ Phys.\ B {\bf 623}, 342 (2002)
  [hep-th/0109154].
  %%CITATION = HEP-TH/0109154;%%
  %235 citations counted in INSPIRE as of 24 Apr 2015
%\cite{Lunin:2002qf}
%\bibitem{Lunin:2002qf} 
  O.~Lunin and S.~D.~Mathur,
  ``Statistical interpretation of Bekenstein entropy for systems with a stretched horizon,''
  Phys.\ Rev.\ Lett.\  {\bf 88}, 211303 (2002)
  [hep-th/0202072].
  %%CITATION = HEP-TH/0202072;%%
  %125 citations counted in INSPIRE as of 24 Apr 2015

\bibitem{firewall-B}
%\bibitem{Braunstein}
S. L. Braunstein, 
``Black hole entropy as entropy of entanglement, 
  or it's curtains for the equivalence principle,''
[arXiv:0907.1190v1 [quant-ph]] 
published as 
%\cite{Braunstein:2009my}
%\bibitem{Braunstein:2009my} 
  S.~L.~Braunstein, S.~Pirandola and K.~Życzkowski,
  ``Better Late than Never: Information Retrieval from Black Holes,''
  Phys.\ Rev.\ Lett.\  {\bf 110}, no. 10, 101301 (2013),
  %[arXiv:0907.1190 [quant-ph]].
  %%CITATION = ARXIV:0907.1190;%%
  %170 citations counted in INSPIRE as of 13 May 2015
  for a similar prediction from different assumptions.

\hide{
\bibitem{WheelerDelayed}
J.~A.~Wheeler, ``Delayed Choice,''
in {\em Mathematical foundations of quantum theory},
Proc. New Orleans Conf. on The mathematical foundations of quantum theory, 
ed. A.R. Marlow (Academic, New York, 1978)
[reprinted in {\em Quantum theory and measurements},
eds. J.~A.~Wheeler and W.~H.~Zurek
(Princeton Univ. Press, Princeton, NJ, 1983) pp. 182-213].

%\cite{Hellmuth:1987zz}
\bibitem{Hellmuth:1987zz}
T.~Hellmuth, H.~Walther, A.~Zajonc and W.~Schleich,
``Delayed-choice experiments in quantum interference,''
Phys. Rev. A \textbf{35}, 2532-2541 (1987)
doi:10.1103/PhysRevA.35.2532
%24 citations counted in INSPIRE as of 05 Sep 2021

%\cite{Kim:1999rj}
\bibitem{Kim:1999rj}
Y.~H.~Kim, R.~Yu, S.~P.~Kulik, Y.~H.~Shih and M.~O.~Scully,
``A delayed choice quantum eraser,''
Phys. Rev. Lett. \textbf{84}, 1-5 (2000)
doi:10.1103/PhysRevLett.84.1
[arXiv:quant-ph/9903047 [quant-ph]].
%59 citations counted in INSPIRE as of 05 Sep 2021

\bibitem{JWGT}
V.~Jacques, E.~Wu, F.~Grosshans, F.~Treussart, 
P.~Grangier, A.~Aspect, and J.-F. ROCH,
``Experimental Realization of Wheeler's Delayed-Choice Gedanken Experiment,''
Science\ {\bf 315}, 966-968 (2007).
}

%\cite{Kawai:2013mda}
\bibitem{Kawai:2013mda} 
  H.~Kawai, Y.~Matsuo and Y.~Yokokura,
  ``A Self-consistent Model of the Black Hole Evaporation,''
  Int.\ J.\ Mod.\ Phys.\ A {\bf 28}, 1350050 (2013)
  [arXiv:1302.4733 [hep-th]].
  %%CITATION = ARXIV:1302.4733;%%
  %8 citations counted in INSPIRE as of 31 Mar 2015

%\cite{Kawai:2014afa}
\bibitem{Kawai:2014afa} 
  H.~Kawai and Y.~Yokokura,
  ``Phenomenological Description of the Interior of the Schwarzschild Black Hole,''
  Int.\ J.\ Mod.\ Phys.\ A {\bf 30}, 1550091 (2015)
  doi:10.1142/S0217751X15500918
  [arXiv:1409.5784 [hep-th]].
  %%CITATION = doi:10.1142/S0217751X15500918;%%
  %29 citations counted in INSPIRE as of 10 Jan 2020
  %\cite{Kawai:2015uya}
  
%\cite{Kawai:2020rmt}
\bibitem{Kawai:2020rmt}
H.~Kawai and Y.~Yokokura,
``Black Hole as a Quantum Field Configuration,''
Universe \textbf{6}, no.6, 77 (2020)
doi:10.3390/universe6060077
[arXiv:2002.10331 [hep-th]].
%3 citations counted in INSPIRE as of 18 Jul 2020

\detail{
%\cite{Penington:2019npb}
\bibitem{Penington:2019npb}
G.~Penington,
``Entanglement Wedge Reconstruction and the Information Paradox,''
JHEP \textbf{09}, 002 (2020)
doi:10.1007/JHEP09(2020)002
[arXiv:1905.08255 [hep-th]].
%137 citations counted in INSPIRE as of 12 Nov 2020

%\cite{Almheiri:2019psf}
\bibitem{Almheiri:2019psf}
A.~Almheiri, N.~Engelhardt, D.~Marolf and H.~Maxfield,
``The entropy of bulk quantum fields and the entanglement wedge of an evaporating black hole,''
JHEP \textbf{12}, 063 (2019)
doi:10.1007/JHEP12(2019)063
[arXiv:1905.08762 [hep-th]].
%138 citations counted in INSPIRE as of 12 Nov 2020

%\cite{Almheiri:2019hni}
\bibitem{Almheiri:2019hni}
A.~Almheiri, R.~Mahajan, J.~Maldacena and Y.~Zhao,
``The Page curve of Hawking radiation from semiclassical geometry,''
JHEP \textbf{03}, 149 (2020)
doi:10.1007/JHEP03(2020)149
[arXiv:1908.10996 [hep-th]].
%260 citations counted in INSPIRE as of 09 Oct 2021

%\cite{Page:1993wv}
\bibitem{Page:1993wv}
D.~N.~Page,
``Information in black hole radiation,''
Phys. Rev. Lett. \textbf{71}, 3743-3746 (1993)
doi:10.1103/PhysRevLett.71.3743
[arXiv:hep-th/9306083 [hep-th]].
%543 citations counted in INSPIRE as of 09 Oct 2021

%\cite{Page:2013dx}
\bibitem{Page:2013dx}
D.~N.~Page,
``Time Dependence of Hawking Radiation Entropy,''
JCAP \textbf{09}, 028 (2013)
doi:10.1088/1475-7516/2013/09/028
[arXiv:1301.4995 [hep-th]].
%150 citations counted in INSPIRE as of 09 Oct 2021
}

\hide{% beginning of hide
%\cite{tHooft:2018zwd}
\bibitem{tHooft:2018zwd}
G.~'t Hooft,
``What happens in a black hole when a particle meets its antipode,''
[arXiv:1804.05744 [gr-qc]].
%5 citations counted in INSPIRE as of 01 Nov 2021

%\cite{Mathur:2020ely}
\bibitem{Mathur:2020ely}
S.~D.~Mathur,
``The VECRO hypothesis,''
doi:10.1142/S0218271820300098
[arXiv:2001.11057 [hep-th]].
%10 citations counted in INSPIRE as of 01 Nov 2021

%\cite{Ho:2021huh}
\bibitem{Ho:2021huh}
P.~M.~Ho,
``Final-State Condition And Dissipative Quantum Mechanics,''
[arXiv:2103.04732 [hep-th]].
%0 citations counted in INSPIRE as of 02 May 2021
}% end of hide

%\cite{Ho:2018jkm}
\bibitem{Ho:2018jkm}
P.~M.~Ho and Y.~Matsuo,
``On the Near-Horizon Geometry of an Evaporating Black Hole,''
JHEP \textbf{07}, 047 (2018)
doi:10.1007/JHEP07(2018)047
[arXiv:1804.04821 [hep-th]].
%10 citations counted in INSPIRE as of 14 Nov 2021

%\cite{Terno:2019kwm}
\bibitem{Terno:2019kwm}
D.~Terno,
``Self-consistent description of a spherically-symmetric gravitational collapse,''
Phys. Rev. D \textbf{100}, no.12, 124025 (2019)
doi:10.1103/PhysRevD.100.124025
[arXiv:1903.04744 [gr-qc]].
%14 citations counted in INSPIRE as of 14 Nov 2021

%\cite{Ho:2019kte}
\bibitem{Ho:2019kte}
P.~M.~Ho and Y.~Matsuo,
``Trapping Horizon and Negative Energy,''
JHEP \textbf{06}, 057 (2019)
doi:10.1007/JHEP06(2019)057
[arXiv:1905.00898 [gr-qc]].
%6 citations counted in INSPIRE as of 14 Nov 2021

\bibitem{ShortDistance}
%\cite{Ho:2019qiu}
P.~M.~Ho, Y.~Matsuo and Y.~Yokokura,
``Distance between collapsing matter and apparent horizon in evaporating black holes,''
Phys. Rev. D \textbf{104}, no.6, 064005 (2021)
doi:10.1103/PhysRevD.104.064005
[arXiv:1912.12863 [gr-qc]].
%7 citations counted in INSPIRE as of 14 Nov 2021

%\cite{Murk:2021cla}
\bibitem{Murk:2021cla}
S.~Murk and D.~R.~Terno,
``Physical black holes in semiclassical gravity,''
[arXiv:2110.12761 [gr-qc]].
%2 citations counted in INSPIRE as of 14 Nov 2021

\end{thebibliography}
\end{document}